\documentclass[aps,reprint,amsmath,amssymb]{revtex4-2}

\usepackage{bm}
\usepackage{amsfonts}

\usepackage{float}
\usepackage{subfigure}
\usepackage{epstopdf}
\usepackage{epsfig}
\usepackage{color,ulem}
\usepackage{multirow}
\usepackage{tabularx}
\usepackage{colortbl}
\usepackage{tabu}
\usepackage{array}

\usepackage{graphicx}
\graphicspath{{graph/},{pics/}}
\usepackage{booktabs} 
\usepackage{longtable}
\usepackage{dcolumn}
\usepackage{bm}
\usepackage[utf8]{inputenc}
\usepackage[T1]{fontenc}
\usepackage{natbib}
\usepackage{booktabs, array, mathptmx, float, tabularx, booktabs, lipsum,multirow}
\usepackage{amsmath}
\usepackage{siunitx, xcolor}
\usepackage[version=4]{mhchem}
\usepackage[colorlinks,linkcolor=blue,anchorcolor=blue,citecolor=blue]{hyperref}

\begin{document}

\title{Quantum anomalous Hall effect with high Chern number in two dimensional ferromagnets Ti$_2$TeSO}

\author{Panjun Feng$^{1,2}$}
\author{Miao Gao$^{3}$}
\author{Xun-Wang Yan$^{4}$}\email{yanxunwang@163.com}
\author{Fengjie Ma$^{1,2}$}\email{fengjie.ma@bnu.edu.cn}

\affiliation{$^{1}$The Center for Advanced Quantum Studies and School of Physics and Astronomy, Beijing Normal University, Beijing 100875, China}
\affiliation{$^{2}$Key Laboratory of Multiscale Spin Physics (Ministry of Education), Beijing Normal University, Beijing 100875, China}
\affiliation{$^{4}$Department of Physics, School of Physical Science and Technology, Ningbo University, Zhejiang 315211, China}
\affiliation{$^{4}$College of Physics and Engineering, Qufu Normal University, Qufu, Shandong 273165, China}

\date{\today}

\begin{abstract}
Two-dimensional Chern insulators have emerged as crucial platforms for the realization of the quantum anomalous Hall effect, and as such have attracted significant interest in spintronics and topological quantum physics due to their unique coexistence of spontaneous magnetization and nontrivial topological characteristics. Nonetheless, substantial challenges persist in such systems, encompassing spin entanglement and the possession of only one edge state (Chern number $C$=1), which significantly hinder their practical applications. Herein, we propose a novel two-dimensional ferromagnetic half-semi-Weyl-metal, monolayer Ti$_2$TeSO, that exhibits exceptional electronic properties. Its majority spin channel possesses only a pair of symmetry-protected Weyl points at the Fermi level, while the states of minority one locate far away from the Fermi level. When spin-orbit coupling is included, a substantial band gap of $\sim$ 92.8 meV is induced at the Weyl points. Remarkably, the emergence of dual dissipationless chiral edge channels and a quantized Hall conductivity plateau at 2$e^2$/$h$ collectively establish monolayer Ti$_2$TeSO as a high-Chern-number insulator with $C$=2. Furthermore, it is demonstrated that valley polarization can be achieved and controlled through the application of strain and the manipulation of the direction of magnetization. The first-principles calculations, in conjunction with Monte Carlo simulations, yield a Curie temperature of 161 K for monolayer Ti$_2$TeSO, thereby indicating the plausibility of coexistence of valley polarization and topological states at elevated temperatures. These findings could provide a foundation for the development of multi-channels dissipationless transport devices and nonvolatile multistate memory architectures.
\end{abstract}

\maketitle

\section{Introduction}

Topological phases of matter have attracted lots of interest in the fields of material sciences and condensed matter physics in recent decades, which have emerged as a fascinating research area, promising to unveil exotic phenomena and to provide new paradigms for the next-generation devices \cite{Wen2017a,Moore2007,ZhangSC2006,Bi2Se3,Weng2015a,Xia2009,Liu2011}. The so-called Chern insulator (CI) or quantum anomalous Hall effect (QAHE), system exhibiting quantized Hall conductance in the absence of an external magnetic field, is of particular interest \cite{Chang2013b,Sha2024,Weng2015b,Bai2025,Yu2010}. 

The CIs are characterized by the topological Chern number $C$, with the quantum Hall conductivity $\sigma_{xy}$ = $Ce^2/h$. Here, $C$ corresponds to the number of the gapless chiral modes residing at the CIs' surface/edge states, $h$ is the Planck constant, and $e$ is represented as the elementary charge \cite{Kane2010, VonKlitzing2005}. Owing to the existence of zero-field dissipationless chiral edge states, the QAHE has great potential to be used for designing high-speed and low-power-consumption spintronic devices \cite{He2018a,Tokura2019,Liang2025}. Furthermore, in order to improve the efficiency of transports, the CIs with a greater number of chiral edge modes, i.e., a higher Chern number $C$, are of significant importance and high demand. 

However, although the fundamental principle of QAHE was theoretically proposed by Haldane in 1988 \cite{Haldane1988b}, its realizations in experiment have only been achieved recently in the thin films of (Bi,Sb)$_2$Te$_3$ family doped by Cr or/and V atoms \cite{Chang2013b,Checkelsky2014a,Mogi2015}, MnBi$_2$Te$_4$ flakes \cite{Deng2020a,Li2019b,Wang2025a}, twisted bilayer graphene \cite{Pierce2021,Geisenhof2021,Serlin2020}, and the transition metal dichalcogenide moir{\'e} superlattices \cite{Zeng2023,Tao2024}. Unfortunately, the process of magnetic ions doping in the (Bi,Sb)$_2$Te$_3$ systems introduces disorder to both the electrical and magnetic properties, thereby reducing the temperature at which a QAHE can be realized \cite{Chang2015, Kudla2019, Wang2018}, while for the other systems the low Curie or N{\'e}el temperatures ($T_c$) fundamentally limit the operating temperature for realizing the QAHE. Moreover, the Chern number of obtained QAHE in these two-dimensional (2D) magnets is usually quantized to $C$= 1, enabling only single dissipationless edge channel where contact resistance imposes severe limitations for integrated circuit interconnects \cite{Wang2013c,Zhao2020a}. These limitations emphasise the critical need to discover 2D CIs that exhibit both high $T_c$ and Chern numbers, which are critical for the practical application of topological quantum devices.

In this work, we propose a 2D ferromagnetic (FM) topological half-semi-metal, the monolayer Ti$_2$TeSO, with a $T_c$ of 161 K. Through the first-principles calculations and tight-binding (TB) model analysis, we find that this material exhibits a quantized Hall conductivity of 2$e^2$/$h$ and clear dual-channel chiral edge states which indicates that the system is a high-Chern number CI. Moreover, the system exhibits significant valley polarization when uniaxial strain is applied or the direction of magnetization is oriented. The effect has been previously reported to be achieved in tetragonal lattices only by means of atomic substitution or electric field control \cite{Guo2024c,Liu2021c,Shen2017,Ren2025}. This work therefore provides a new strategy that not only achieves but also enables dynamic flipping of valley polarization, thus establishing an ideal platform for the design of topological quantum devices based on valley degrees of freedom.

\section{Computational method}

In our studies, the structural optimization and electronic properties calculations of monolayer Ti$_2$TeSO were carried out with the Quantum ESPRESSO package \cite{Giannozzi2009, Giannozzi2017}, in which the generalized gradient approximation (GGA) of the Perdew-Burke-Ernzerhof formula for the exchange-correlation potentials and the ultrasoft pseudopotentials were adopted \cite{Perdew1996}. A corrective Hubbard-like U term was introduced to treat the strong on-site Coulomb interaction of the localized electrons of the transition metal ions \cite{Cococcioni2005, Liechtenstein1995}. The effective values of U used in calculations was 3.0 eV for Ti-3$d$  electrons. The plane-wave kinetic-energy cutoff and the energy cutoff for charge density were set as 85 and 680 Ry, respectively. A mesh of \mbox{24$\times$24$\times$1} k-points grid was used for sampling the Brillouin zone, and the Methfessel-Paxton first-order spreading technique was adopted \cite{Methfessel1989}. In order to avoid the residual interaction between adjacent layers, a 20 \AA{} vacuum layer was used. During the simulations, all structural geometries were fully optimized to achieve the minimum energy. Phonon band dispersions were calculated by using density functional perturbation theory based on the PHONOPY program \cite{Togo2015}. In the \textit{ab initio} molecular dynamics simulations, a 3$\times$3$\times$1 supercell was employed and the temperature was kept at 300 K for 5 ps with a time step of 1 fs in the canonical ensemble. The $T_c$ is evaluated by the Monte Carlo method enclosed in the software package MTC \cite{Zhang2021}. The edge states were studied using TB methods by a combination of Wannier90 and WannierTools software packages \cite{Mostofi2008,Wu2018}.

\section{Results and disscussions}

\subsection{Atomic structure and stability} 

\begin{figure}
	\begin{center}
		\includegraphics[width=1.0\columnwidth]{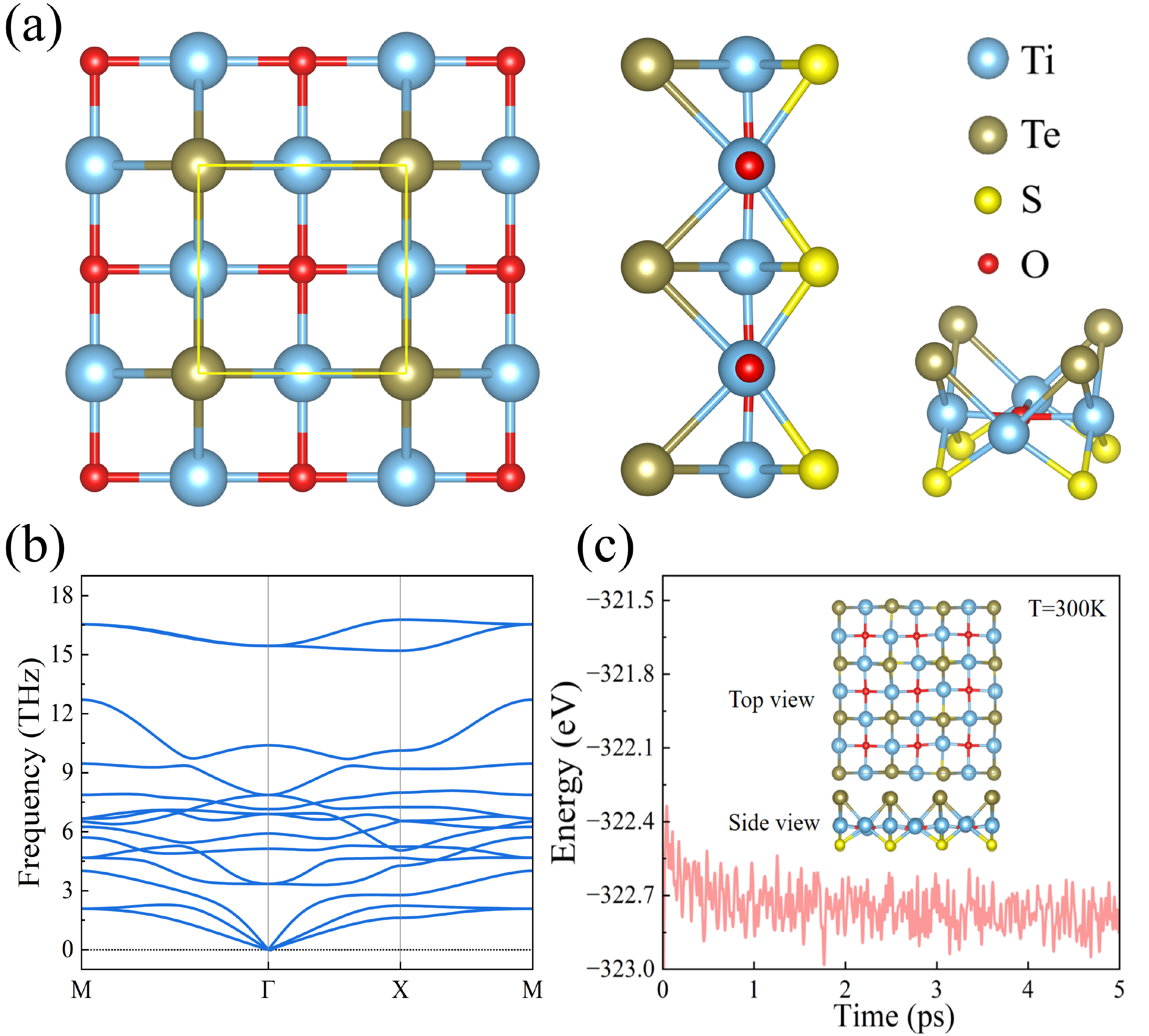}
		\caption{(a) Top view and side view of atomic structure of Ti$_2$TeSO. (b) The phonon spectrum of monolayer Ti$_2$TeSO. (c) The total energy evolutions  with respect to time in molecular dynamics simulations.
		} \label{structmodel}
	\end{center}
\end{figure}

As demonstrated in Fig.\ref{structmodel}(a), the monolayer Ti$_2$TeSO adopts a square lattice structure, characterized by a  space group of $P4mm$ (No. 99), which is affiliated to the C$_{4v}$ point group symmetry. This material is of a particular variety of Janus materials with a spontaneous spatial inversion symmetry breaking. Within each unit cell, there are five atoms, with the Ti-O layer positioned between the Te layer and the S layer. 

\begin{table}[b]
    \begin{center}                                   
    \caption{The relative energies (in unit of meV) of monolayer Ti$_2$TeSO for different possible magnetic states, in which the energy of FM state is set to zero.}
    \label{T1}
    \begin{ruledtabular}
    \begin{tabular}{cccccc}
    \textrm{ }&
    \textrm{FM}&
    \textrm{NM}&
    \textrm{Neel}&
    \textrm{Stripe}&
    \textrm{Zigzag} \\
    \colrule
   Energy/meV & 0   & 461.58 & 161.53 & 79.19 & 122.33    \\ 
    \end{tabular}
    \end{ruledtabular}
    \end{center}
\end{table}

Given the presence of 3$d$ transition metal atoms within the compound, multiple possible magnetic ground states are considered, including stripe-antiferromagnetic (AFM), Neel-AFM, FM, zigzag-AFM, and non-magnetic (NM) states. Our calculations reveal that monolayer Ti$_2$TeSO has a FM ground state, resulting in a magnetic space group of \textit{P4m$\prime$m$\prime$} (No. 99.5.827) \cite{findsym,Stokes_FINDSYM}. The relative energies for these different magnetic states are enumerated in Table \ref{T1}. It is evident that the FM state possesses a significantly lower energy in comparison to the other states. The optimized lattice constants of monolayer Ti$_2$TeSO are found to be about \mbox{$a$ = $b$ = 4.15 \AA}, and the magnetic moments are mainly localized around each Ti atom with a magnitude of approximately 0.8 $\mu_B$.

The dynamical and thermal stability of monolayer Ti$_2$TeSO both have been checked by the calculations of phonon spectrum and \textit{ab initio} molecular dynamic simulations \cite{Giannozzi2009, Giannozzi2017, Togo2015}. The absence of imaginary frequency modes in the phonon spectrum, and well maintained crystal structure and steadily potential energy evolution curve up to a temperature of 300 K all indicate the good dynamical and thermal stability of monolayer Ti$_2$TeSO, as shown in Figs. \ref{structmodel}(b-c).

\subsection{Electronic and topological properties} 

\begin{figure}
    \begin{center}
    \includegraphics[width=1.0\columnwidth]{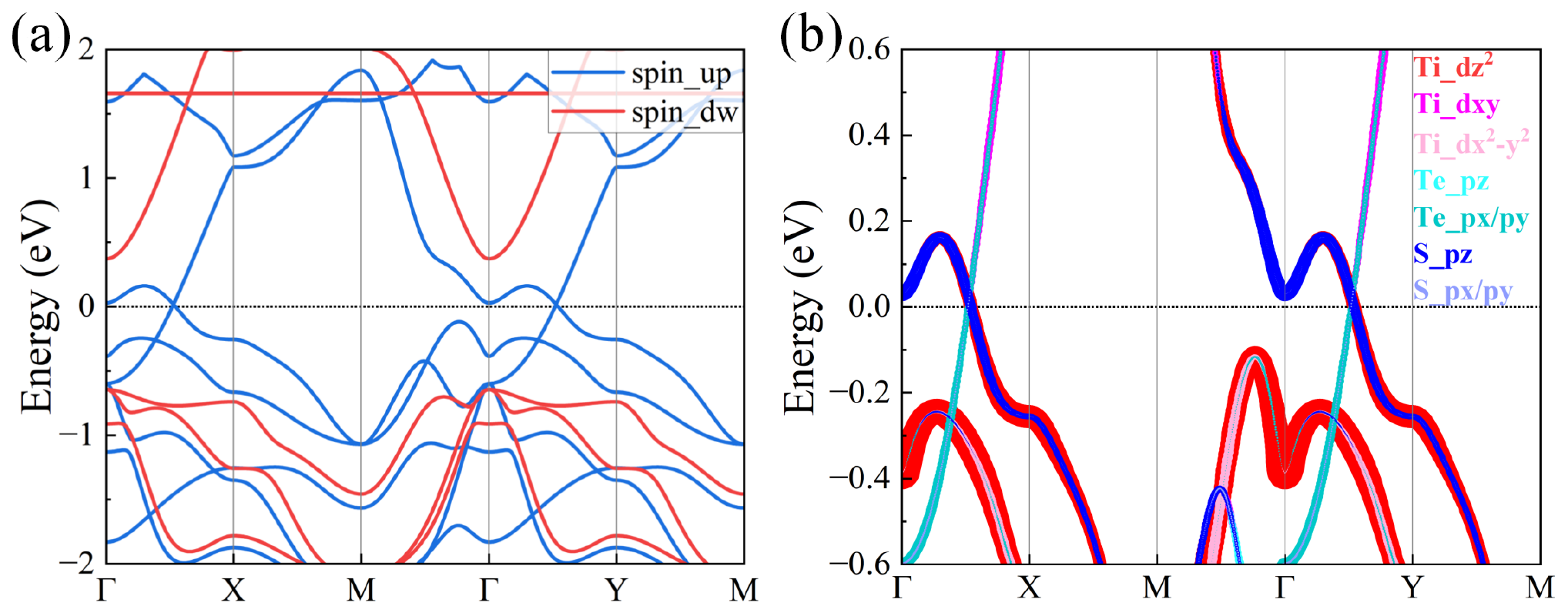}
    \caption{(a) The band structure of monolayer Ti$_2$TeSO without SOC. (b) The Ti-$d$, Te-$p$, and S-$p$ orbital projection of the bands near the Fermi level. 
    } \label{band}
    \end{center}
\end{figure}

Figure \ref{band}(a) gives the calculated electronic band structure of 2D monolayer Ti$_2$TeSO. In the absence of SOC, the material is a perfect half-semi-metal, whose majority spin exhibits a linearly-crossing band dispersion and vanishing density of states at the Fermi level while the minority one has a moderate, $\sim$ 1 eV, insulating band gap. A full (100$\%$) spin polarization near the Fermi level is found in the monolayer Ti$_2$TeSO. Since the two intersecting bands of the majority spin belong to inequivalent irreducible representations, A$^\prime$ and A$^{\prime\prime}$ of the $C_s$ little group respectively, their hybridization is prohibited and the crossing keeps gapless. The material therefore features a perfect gapless half-semi-Weyl-metal nature with linear band crossings exactly located at the Fermi level. As elucidated by the orbital projection band structures, shown in Fig.\ref{band}(b), the states near the Fermi level are predominantly constituted by the hybridization of the $d_{z^{2}}$ and $d_{xy}$ orbitals of Ti atoms, the $p_x$/$p_y$ orbitals of Te atoms, and the $p_z$ orbitals of S atoms. This implies that monolayer Ti$_2$TeSO has a non-negligible SOC effect near the Fermi level.

\begin{figure}
    \begin{center}
    \includegraphics[width=1.0\columnwidth]{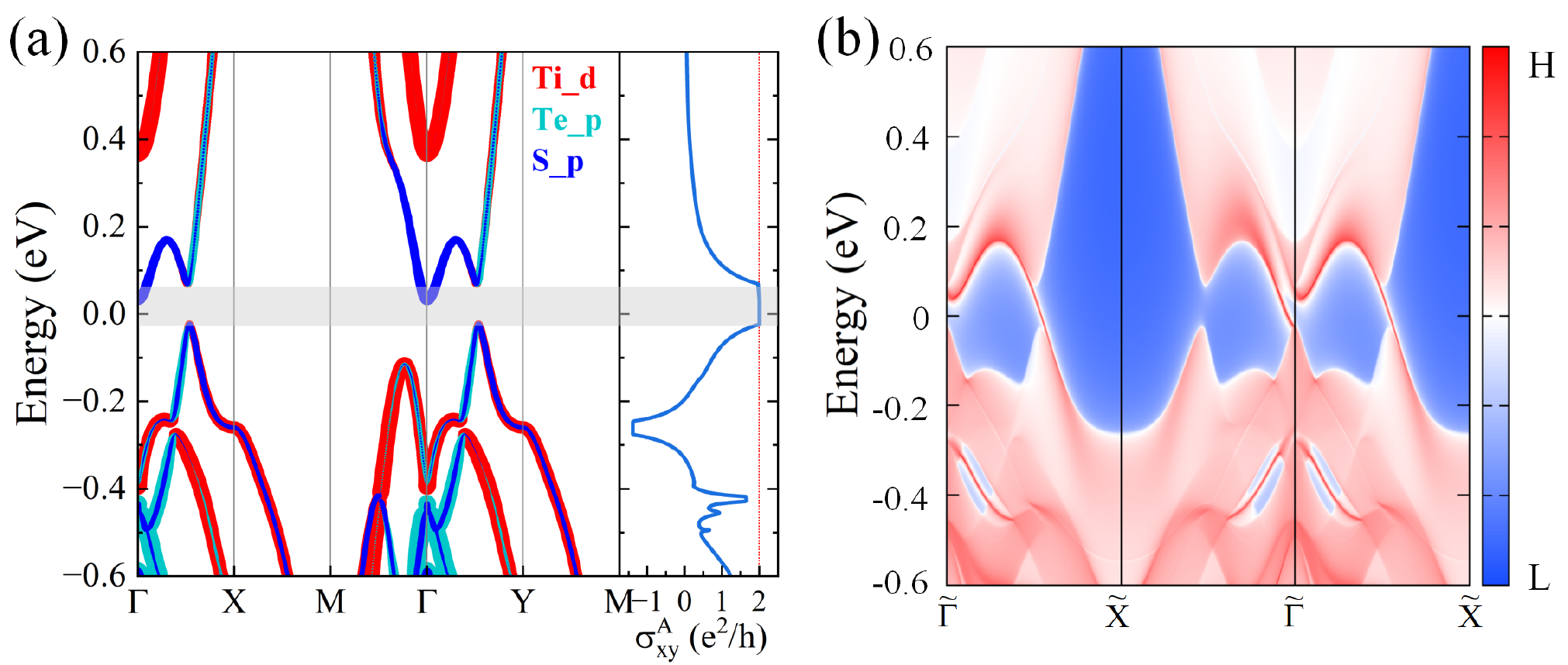}
    \caption{(a) The orbital projection of the bands and AHC of monolayer Ti$_2$TeSO with SOC. (b) The edge states of monolayer Ti$_2$TeSO cut along the [010] direction.
    } \label{topo}
    \end{center}
\end{figure}

Once SOC is introduced, energy gaps of $\sim$92.8 meV are opened at the Weyl points, turning the linear band dispersions into quadratic ones, accompanied by an inversion of the energy bands, as shown in the left panel of Fig. \ref{topo}(a). This mechanism is driven by the heavy-element characteristics of Te atoms and the synergistic effect of $d$-$p$ hybridization between Ti and Te/S atoms. The energy bands near the Fermi level still remain spin-polarized, which can provide fully spin-resolved currents, offering great promise for spintronic applications. 

As topological phase transitions may occur during the transition of the band gap from a closed to an open state, we constructed the maximum localization Wannier functions to obtain an effective TB Hamiltonian using the Wannier90 program, thus further investigating the topological properties of monolayer Ti$_2$TeSO \cite{Mostofi2008, Wu2018}. The anomalous Hall conductivity (AHC) and edge states were calculated. As demonstrated in the right panel of Fig.\ref{topo}(a), the AHC exhibits a plateau of 2$e^2$/$h$ in proximity to the Fermi level, with the width of this plateau being equivalent to the size of the gaps that have been opened. This finding suggests the presence of a QAHE with a high Chern number ($C$=2). As illustrated in Fig.\ref{topo}(b), the edge states along the [010] direction substantiate the existence of two gapless edge channels connecting the conduction and valence bands within the energy gap. This further corroborates the assertion that the monolayer Ti$_2$TeSO is a high-Chern-number CI, whose topological properties significantly differ from those of traditional $C$=1 QAHE systems, thus providing a new platform for exploring multi-channels quantum transports.

\begin{figure}
    \begin{center}
    \includegraphics[width=1.0\columnwidth]{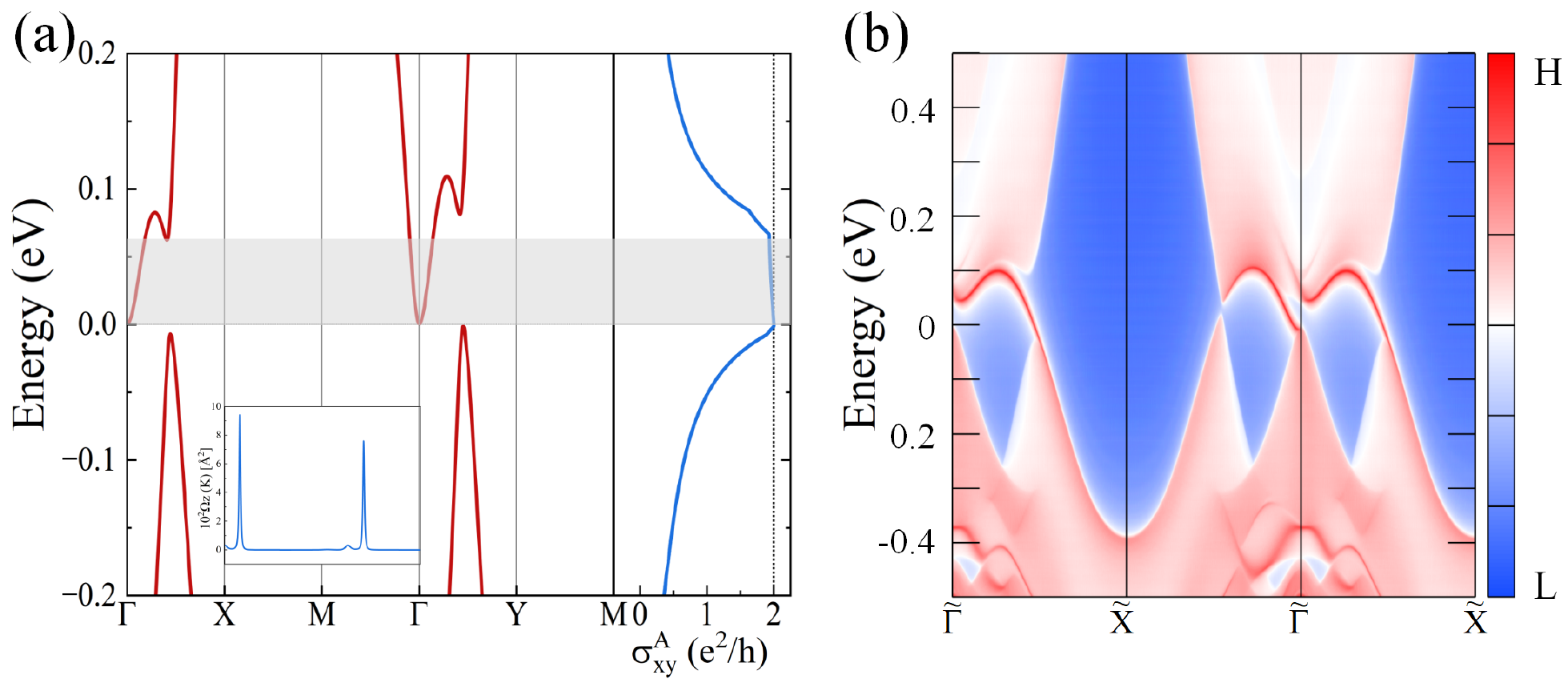}
    \caption{The band structure and AHC (a) and the edge states cut along the [010] direction (b) of monolayer Ti$_2$TeSO with SOC under 1$\%$ uniaxial strain along the $x$-axis.
    } \label{strain}
    \end{center}
\end{figure}

The topological properties of monolayer Ti$_2$TeSO remain robust against external strain. Figure \ref{strain}(a) illustrates the band structure and AHC of monolayer Ti$_2$TeSO with SOC under 1$\%$ uniaxial strain along the $x$-axis. Such strain induces a breaking of the C$_{4z}$ symmetry that connects the two valleys, thereby inducing valley polarization. The presence of a constant positive Berry curvature throughout the Brillouin zone serves to confirm the topologically non-trivial nature of the system. As shown in Fig. \ref{strain}, the AHC manifests a quantized plateau of 2$e^2$/$h$, accompanied by two discrete edge states bridging the conduction and valence bands. These features indicates the coexistence of the high Chern number QAHE and the anomalous valley Hall effect. Moreover, in the presence of more appropriate uniaxial or biaxial strain, it is anticipated that the system will undergo a topological phase transition from $C$=2 to $C$=1 or $C$=0, when the lowest conduction band and highest valence band intersected at one or both valleys, removing the band inversion accompanied by a re-open of energy gap due to the lack of symmetry protection. 

\begin{figure}
    \begin{center}
    \includegraphics[width=1.0\columnwidth]{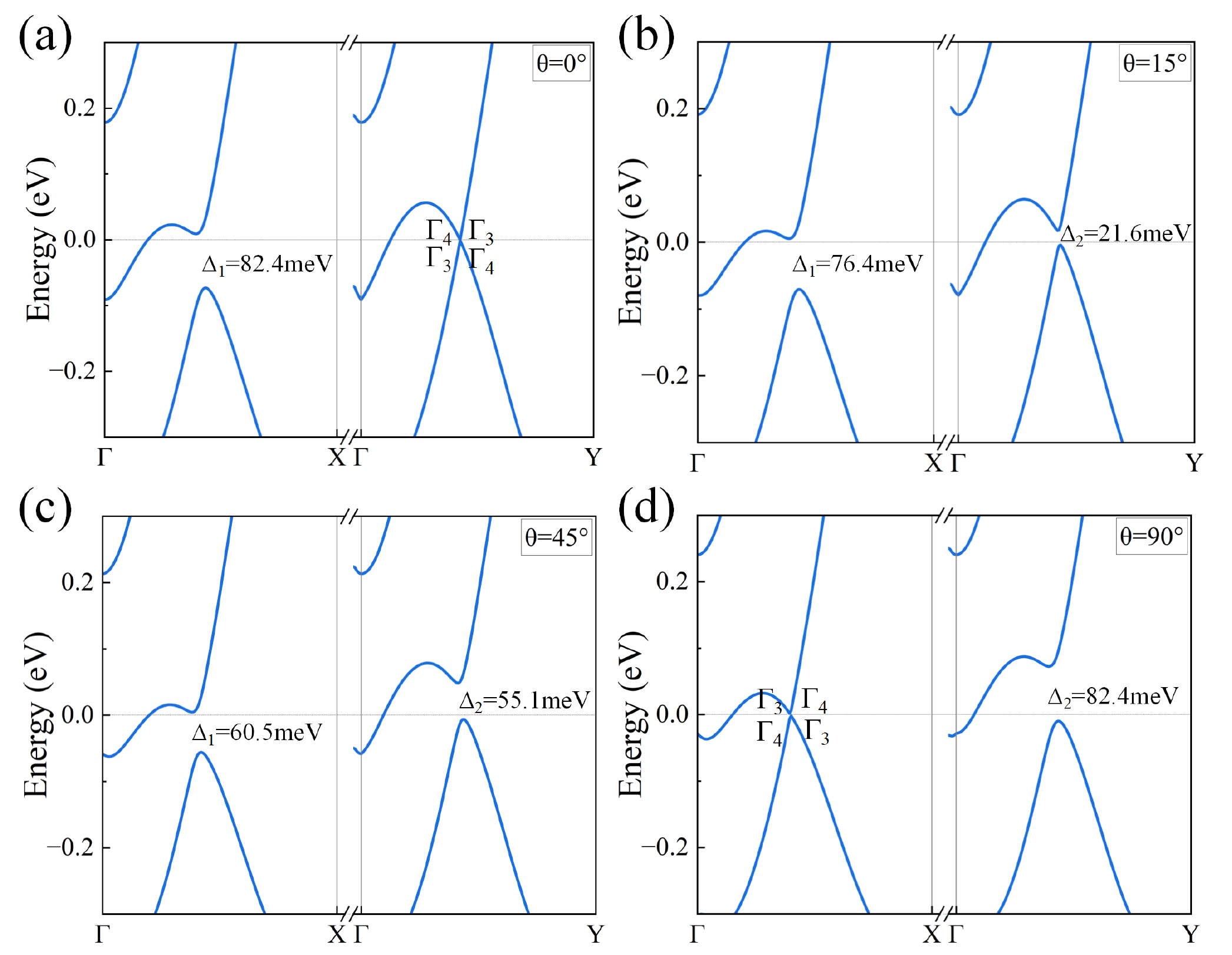}
    \caption{(a-d) The variation of band structures of monolayer Ti$_2$TeSO as the angle between the magnetization direction and the $x$-axis ($\theta$) is increased. The SOC effect is considered. 
    } \label{Minplane}
    \end{center}
\end{figure}

In addition, the valley polarization of monolayer Ti$_2$TeSO could be also introduced by orienting the direction of magnetization. As illustrated in Fig. \ref{Minplane}, the modulation of band structures and valley polarization is contingent on the direction of magnetization, with the angle relative to the $x$-axis denoted by the variable $\theta$. In the circumstance that the direction of magnetization is aligned along the $x$-axis ($\theta$ = 0$^\circ$), the Weyl point located on the $\Gamma$-Y path still exists, since the highest valence band and lowest conduction band belong to two different irreducible representations, $\Gamma_3$ and $\Gamma_4$ of $C_s$ double point group respectively, thereby engendering a topologically protected linear crossing. In contrast, the other Weyl node located on the $\Gamma$-X path is not symmetry-protected and a band gap of about 82.4 meV is developed. As $\theta$ increases, both of the Weyl nodes simultaneously develop energy gaps, in which we label the gaps along the $\Gamma$-X and $\Gamma$-X paths as $\Delta_1$ and $\Delta_2$, respectively.  $\Delta_1$ gradually decreases while $\Delta_2$ correspondingly increases, demonstrating an antagonistic regulation behavior. When $\theta$ = 90{$^{\circ}$}, $\Delta_1$ reaches its maximum value and $\Delta_2$ goes to zero, leading to a mirror band structure in comparison with the one of $\theta$ = 0$^\circ$ configuration, thereby achieving a controllable reversal of valley polarization direction. Therefore, the direction-of-magnetization-driven topological phase transition enables dynamic switching between half-semi-Weyl-metal and valley polarized insulator states, providing a new degree of freedom for designing spin-valley-coupled multifunctional devices.

\subsection{Heisenberg Model}

Since the critical parameters, such as the $T_c$, is necessary for evaluating the FM materials for their possible technical applications. Monte Carlo simulations based on the 2D Heisenberg model were therefore employed to estimate the $T_c$ of monolayer Ti$_2$TeSO. The Hamiltonian is defined as follows:
\begin{align}
H & = J_1\sum_{\left \langle ij \right \rangle}S_{i}S_{j}+J_2\sum_{\ll ij \gg }S_{i}S_{j}+A\sum_{i}{\left ( S_{iz} \right )}^2    
\end{align} 
where $J_1$ and $J_2$ are the nearest and the next nearest neighbor exchange parameter, respectively. $S_i$ and $S_j$ denote the spin of the Ti atom in the $i$th and $j$th site, respectively. $A$ represents the unit magnetic anisotropy energy. The magnetic exchange interactions were derived from the energy difference between the AFM and FM ground states calculated using the PBE+U method:
\begin{align}
    J_{1}  & = (E_{FM}-E_{Neel})/8 \\
    J_{2}  & = (E_{FM} +E_{Neel}-2E_{Stripe})/16
\end{align}
where $E_{FM}$, $E_{Neel}$, and $E_{Stripe}$ are the energies of FM, Neel-AFM, and stripe-AFM states per formula cell.

\begin{figure}
    \begin{center}
    \includegraphics[width=1.0\columnwidth]{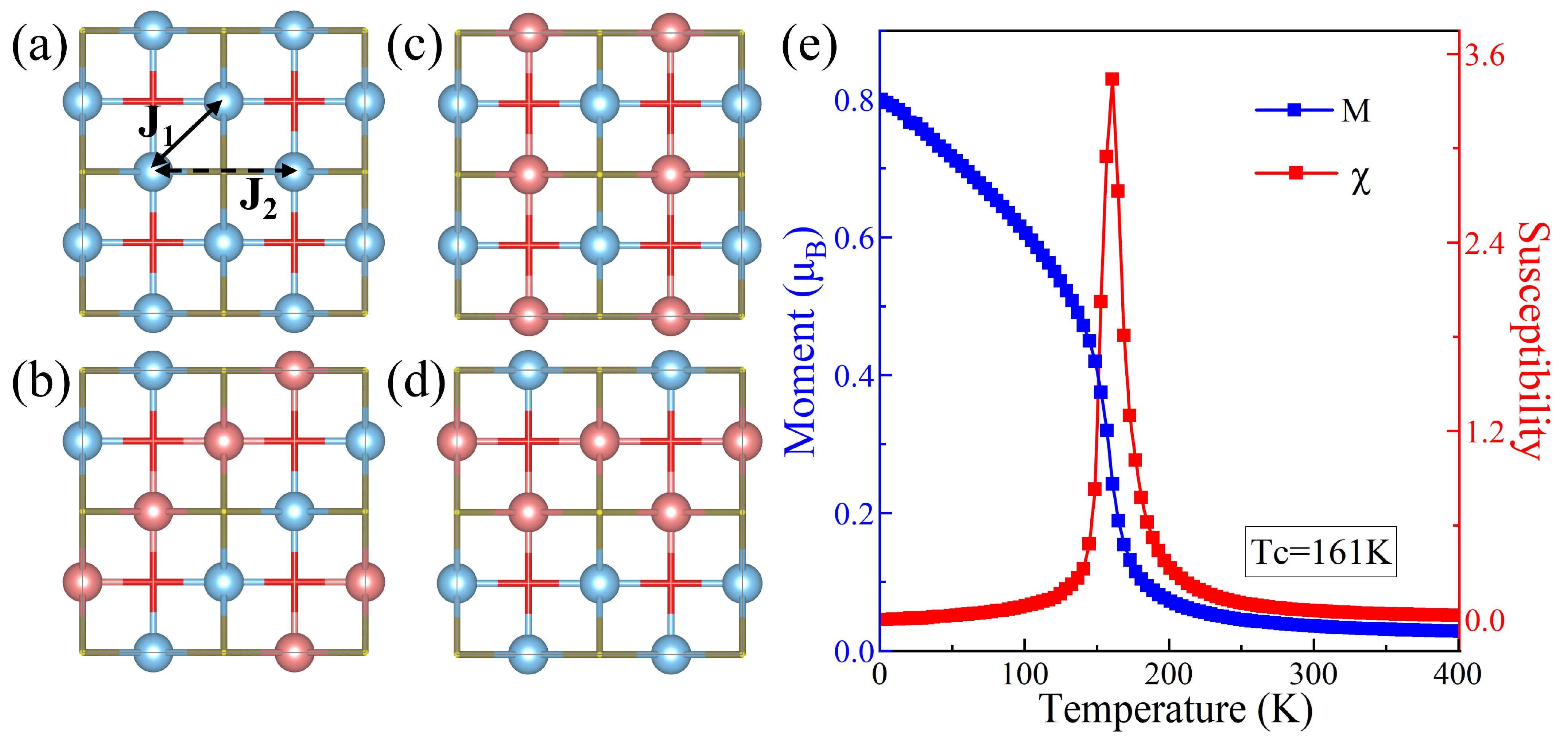}
    \caption{The (a)-(d) plot shows the different magnetic orders of Ti$_2$TeSO, in which the $J_1$ and $J_2$ are the exchange interaction of the nearest neighbor sites and the nearest neighbor sites. (e) Average magnetic moment M and magnetic susceptibility $\chi$ as functions of temperature from the solution of the Heisenberg model for the Ti$_2$TeSO.
    } \label{Tc}
    \end{center}
\end{figure}

For the monolayer Ti$_2$TeSO, the exchange coupling constants are given by $J_1$ = -20.19 meV/S$^2$ and $J_2$ = 0.26 meV/S$^2$. The unit magnetic anisotropy energy is $A$ = 1.04 meV/S$^2$. The spontaneous magnetic susceptibility ($\chi$) with respect to temperature are plotted in Fig. \ref{Tc}(e), and the result shows that the $T_c$ of the Ti$_2$TeSO monolayer is about 161 K, much higher than the temperatures reported that QAHE is realized.

\section{Conclusions}

In summary, a 2D Janus FM half-semi-Weyl-metal, monolayer Ti$_2$TeSO, is proposed as a high-Chern-number CI with integrated topological and valleytronic functionalities. When considering SOC effect, a significant band gap of 92.8 meV is opened at the Weyl points. The first-principles calculations reveal a high Chern number $C$=2, evidenced by a quantized Hall conductivity plateau at 2$e^2/h$ and dual dissipationless chiral edge states. Moreover, the system exhibits magnetization-direction-dependent valley polarization within a tetragonal lattice. Through in-plane magnetization rotation, isolated valley states can be dynamically modulated, enabling valley polarization reversal without the need of external electronic field. Furthermore, Monte Carlo simulations predict a $T_c$ of 161 K, approaching the liquid nitrogen temperature regime, thereby demonstrating the feasibility of coexisting valley polarization and topological states at elevated temperatures. Therefore, the synergistic coexistence of a high Chern number,  tunable valley polarization, and enhanced $T_c$ positions Ti$_2$TeSO as a versatile platform for multi-channels quantum transport and nonvolatile valleytronic devices. This work not only contributes to the design of high Chern number and high-$T_c$ QAHE systems, but also provides a potential candidate for the coexistence of topological states and valley degrees of freedom in non-hexagonal lattices.

\begin{acknowledgments}

This work was financially supported by the National Natural Science Foundation of China under Grants No. 12074040, No. 11974207, No. 12274255, and No. 11974194. F. Ma was also supported by the BNU Tang Scholar. The computations were supported by the Center for Advanced Quantum Studies, Beijing Normal University, and Hongzhiwei Technology.

\end{acknowledgments}

\bibliography{reference.bib}

\begin{thebibliography}{49}%
\makeatletter
\providecommand \@ifxundefined [1]{%
 \@ifx{#1\undefined}
}%
\providecommand \@ifnum [1]{%
 \ifnum #1\expandafter \@firstoftwo
 \else \expandafter \@secondoftwo
 \fi
}%
\providecommand \@ifx [1]{%
 \ifx #1\expandafter \@firstoftwo
 \else \expandafter \@secondoftwo
 \fi
}%
\providecommand \natexlab [1]{#1}%
\providecommand \enquote  [1]{``#1''}%
\providecommand \bibnamefont  [1]{#1}%
\providecommand \bibfnamefont [1]{#1}%
\providecommand \citenamefont [1]{#1}%
\providecommand \href@noop [0]{\@secondoftwo}%
\providecommand \href [0]{\begingroup \@sanitize@url \@href}%
\providecommand \@href[1]{\@@startlink{#1}\@@href}%
\providecommand \@@href[1]{\endgroup#1\@@endlink}%
\providecommand \@sanitize@url [0]{\catcode `\\12\catcode `\$12\catcode `\&12\catcode `\#12\catcode `\^12\catcode `\_12\catcode `\%12\relax}%
\providecommand \@@startlink[1]{}%
\providecommand \@@endlink[0]{}%
\providecommand \url  [0]{\begingroup\@sanitize@url \@url }%
\providecommand \@url [1]{\endgroup\@href {#1}{\urlprefix }}%
\providecommand \urlprefix  [0]{URL }%
\providecommand \Eprint [0]{\href }%
\providecommand \doibase [0]{https://doi.org/}%
\providecommand \selectlanguage [0]{\@gobble}%
\providecommand \bibinfo  [0]{\@secondoftwo}%
\providecommand \bibfield  [0]{\@secondoftwo}%
\providecommand \translation [1]{[#1]}%
\providecommand \BibitemOpen [0]{}%
\providecommand \bibitemStop [0]{}%
\providecommand \bibitemNoStop [0]{.\EOS\space}%
\providecommand \EOS [0]{\spacefactor3000\relax}%
\providecommand \BibitemShut  [1]{\csname bibitem#1\endcsname}%
\let\auto@bib@innerbib\@empty
\bibitem [{\citenamefont {Wen}(2017)}]{Wen2017a}%
  \BibitemOpen
  \bibfield  {author} {\bibinfo {author} {\bibfnamefont {X.-G.}\ \bibnamefont {Wen}},\ }\bibfield  {title} {\bibinfo {title} {{\emph{Colloquium}} : Zoo of quantum-topological phases of matter},\ }\href {https://doi.org/10.1103/RevModPhys.89.041004} {\bibfield  {journal} {\bibinfo  {journal} {Reviews of Modern Physics}\ }\textbf {\bibinfo {volume} {89}},\ \bibinfo {pages} {041004} (\bibinfo {year} {2017})}\BibitemShut {NoStop}%
\bibitem [{\citenamefont {Moore}\ and\ \citenamefont {Balents}(2007)}]{Moore2007}%
  \BibitemOpen
  \bibfield  {author} {\bibinfo {author} {\bibfnamefont {J.~E.}\ \bibnamefont {Moore}}\ and\ \bibinfo {author} {\bibfnamefont {L.}~\bibnamefont {Balents}},\ }\bibfield  {title} {\bibinfo {title} {Topological invariants of time-reversal-invariant band structures},\ }\href {https://doi.org/10.1103/PhysRevB.75.121306} {\bibfield  {journal} {\bibinfo  {journal} {Physical Review B}\ }\textbf {\bibinfo {volume} {75}},\ \bibinfo {pages} {121306} (\bibinfo {year} {2007})}\BibitemShut {NoStop}%
\bibitem [{\citenamefont {Bernevig}\ and\ \citenamefont {Zhang}(2006)}]{ZhangSC2006}%
  \BibitemOpen
  \bibfield  {author} {\bibinfo {author} {\bibfnamefont {B.~A.}\ \bibnamefont {Bernevig}}\ and\ \bibinfo {author} {\bibfnamefont {S.-C.}\ \bibnamefont {Zhang}},\ }\bibfield  {title} {\bibinfo {title} {Quantum spin hall effect},\ }\href {https://doi.org/10.1103/PhysRevLett.96.106802} {\bibfield  {journal} {\bibinfo  {journal} {Phys. Rev. Lett.}\ }\textbf {\bibinfo {volume} {96}},\ \bibinfo {pages} {106802} (\bibinfo {year} {2006})}\BibitemShut {NoStop}%
\bibitem [{\citenamefont {Zhang}\ \emph {et~al.}(2009)\citenamefont {Zhang}, \citenamefont {Liu}, \citenamefont {Qi}, \citenamefont {Dai}, \citenamefont {Fang},\ and\ \citenamefont {Zhang}}]{Bi2Se3}%
  \BibitemOpen
  \bibfield  {author} {\bibinfo {author} {\bibfnamefont {H.}~\bibnamefont {Zhang}}, \bibinfo {author} {\bibfnamefont {C.-X.}\ \bibnamefont {Liu}}, \bibinfo {author} {\bibfnamefont {X.-L.}\ \bibnamefont {Qi}}, \bibinfo {author} {\bibfnamefont {X.}~\bibnamefont {Dai}}, \bibinfo {author} {\bibfnamefont {Z.}~\bibnamefont {Fang}},\ and\ \bibinfo {author} {\bibfnamefont {S.-C.}\ \bibnamefont {Zhang}},\ }\bibfield  {title} {\bibinfo {title} {Topological insulators in bi{\textsubscript{2}}se{\textsubscript{3}}, bi{\textsubscript{2}}te{\textsubscript{3}} and sb{\textsubscript{2}}te{\textsubscript{3}} with a single dirac cone on the surface},\ }\href {https://doi.org/10.1038/nphys1270} {\bibfield  {journal} {\bibinfo  {journal} {Nature Physics}\ }\textbf {\bibinfo {volume} {5}},\ \bibinfo {pages} {438} (\bibinfo {year} {2009})}\BibitemShut {NoStop}%
\bibitem [{\citenamefont {Weng}\ \emph {et~al.}(2015{\natexlab{a}})\citenamefont {Weng}, \citenamefont {Fang}, \citenamefont {Fang}, \citenamefont {Bernevig},\ and\ \citenamefont {Dai}}]{Weng2015a}%
  \BibitemOpen
  \bibfield  {author} {\bibinfo {author} {\bibfnamefont {H.}~\bibnamefont {Weng}}, \bibinfo {author} {\bibfnamefont {C.}~\bibnamefont {Fang}}, \bibinfo {author} {\bibfnamefont {Z.}~\bibnamefont {Fang}}, \bibinfo {author} {\bibfnamefont {B.~A.}\ \bibnamefont {Bernevig}},\ and\ \bibinfo {author} {\bibfnamefont {X.}~\bibnamefont {Dai}},\ }\bibfield  {title} {\bibinfo {title} {Weyl semimetal phase in noncentrosymmetric transition-metal monophosphides},\ }\href {https://doi.org/10.1103/PhysRevX.5.011029} {\bibfield  {journal} {\bibinfo  {journal} {Physical Review X}\ }\textbf {\bibinfo {volume} {5}},\ \bibinfo {pages} {011029} (\bibinfo {year} {2015}{\natexlab{a}})}\BibitemShut {NoStop}%
\bibitem [{\citenamefont {Xia}\ \emph {et~al.}(2009)\citenamefont {Xia}, \citenamefont {Qian}, \citenamefont {Hsieh}, \citenamefont {Wray}, \citenamefont {Pal}, \citenamefont {Lin}, \citenamefont {Bansil}, \citenamefont {Grauer}, \citenamefont {Hor}, \citenamefont {Cava},\ and\ \citenamefont {Hasan}}]{Xia2009}%
  \BibitemOpen
  \bibfield  {author} {\bibinfo {author} {\bibfnamefont {Y.}~\bibnamefont {Xia}}, \bibinfo {author} {\bibfnamefont {D.}~\bibnamefont {Qian}}, \bibinfo {author} {\bibfnamefont {D.}~\bibnamefont {Hsieh}}, \bibinfo {author} {\bibfnamefont {L.}~\bibnamefont {Wray}}, \bibinfo {author} {\bibfnamefont {A.}~\bibnamefont {Pal}}, \bibinfo {author} {\bibfnamefont {H.}~\bibnamefont {Lin}}, \bibinfo {author} {\bibfnamefont {A.}~\bibnamefont {Bansil}}, \bibinfo {author} {\bibfnamefont {D.}~\bibnamefont {Grauer}}, \bibinfo {author} {\bibfnamefont {Y.~S.}\ \bibnamefont {Hor}}, \bibinfo {author} {\bibfnamefont {R.~J.}\ \bibnamefont {Cava}},\ and\ \bibinfo {author} {\bibfnamefont {M.~Z.}\ \bibnamefont {Hasan}},\ }\bibfield  {title} {\bibinfo {title} {Observation of a large-gap topological-insulator class with a single dirac cone on the surface},\ }\href {https://doi.org/10.1038/nphys1274} {\bibfield  {journal} {\bibinfo  {journal} {Nature Physics}\ }\textbf {\bibinfo {volume} {5}},\ \bibinfo {pages} {398} (\bibinfo {year}
  {2009})}\BibitemShut {NoStop}%
\bibitem [{\citenamefont {Liu}\ \emph {et~al.}(2011)\citenamefont {Liu}, \citenamefont {Feng},\ and\ \citenamefont {Yao}}]{Liu2011}%
  \BibitemOpen
  \bibfield  {author} {\bibinfo {author} {\bibfnamefont {C.-C.}\ \bibnamefont {Liu}}, \bibinfo {author} {\bibfnamefont {W.}~\bibnamefont {Feng}},\ and\ \bibinfo {author} {\bibfnamefont {Y.}~\bibnamefont {Yao}},\ }\bibfield  {title} {\bibinfo {title} {Quantum spin hall effect in silicene and two-dimensional germanium},\ }\href {https://doi.org/10.1103/PhysRevLett.107.076802} {\bibfield  {journal} {\bibinfo  {journal} {Physical Review Letters}\ }\textbf {\bibinfo {volume} {107}},\ \bibinfo {pages} {076802} (\bibinfo {year} {2011})}\BibitemShut {NoStop}%
\bibitem [{\citenamefont {Chang}\ \emph {et~al.}(2013)\citenamefont {Chang}, \citenamefont {Zhang}, \citenamefont {Feng}, \citenamefont {Shen}, \citenamefont {Zhang}, \citenamefont {Guo}, \citenamefont {Li}, \citenamefont {Ou}, \citenamefont {Wei}, \citenamefont {Wang}, \citenamefont {Ji}, \citenamefont {Feng}, \citenamefont {Ji}, \citenamefont {Chen}, \citenamefont {Jia}, \citenamefont {Dai}, \citenamefont {Fang}, \citenamefont {Zhang}, \citenamefont {He}, \citenamefont {Wang}, \citenamefont {Lu}, \citenamefont {Ma},\ and\ \citenamefont {Xue}}]{Chang2013b}%
  \BibitemOpen
  \bibfield  {author} {\bibinfo {author} {\bibfnamefont {C.-Z.}\ \bibnamefont {Chang}}, \bibinfo {author} {\bibfnamefont {J.}~\bibnamefont {Zhang}}, \bibinfo {author} {\bibfnamefont {X.}~\bibnamefont {Feng}}, \bibinfo {author} {\bibfnamefont {J.}~\bibnamefont {Shen}}, \bibinfo {author} {\bibfnamefont {Z.}~\bibnamefont {Zhang}}, \bibinfo {author} {\bibfnamefont {M.}~\bibnamefont {Guo}}, \bibinfo {author} {\bibfnamefont {K.}~\bibnamefont {Li}}, \bibinfo {author} {\bibfnamefont {Y.}~\bibnamefont {Ou}}, \bibinfo {author} {\bibfnamefont {P.}~\bibnamefont {Wei}}, \bibinfo {author} {\bibfnamefont {L.-L.}\ \bibnamefont {Wang}}, \bibinfo {author} {\bibfnamefont {Z.-Q.}\ \bibnamefont {Ji}}, \bibinfo {author} {\bibfnamefont {Y.}~\bibnamefont {Feng}}, \bibinfo {author} {\bibfnamefont {S.}~\bibnamefont {Ji}}, \bibinfo {author} {\bibfnamefont {X.}~\bibnamefont {Chen}}, \bibinfo {author} {\bibfnamefont {J.}~\bibnamefont {Jia}}, \bibinfo {author} {\bibfnamefont {X.}~\bibnamefont {Dai}}, \bibinfo {author} {\bibfnamefont
  {Z.}~\bibnamefont {Fang}}, \bibinfo {author} {\bibfnamefont {S.-C.}\ \bibnamefont {Zhang}}, \bibinfo {author} {\bibfnamefont {K.}~\bibnamefont {He}}, \bibinfo {author} {\bibfnamefont {Y.}~\bibnamefont {Wang}}, \bibinfo {author} {\bibfnamefont {L.}~\bibnamefont {Lu}}, \bibinfo {author} {\bibfnamefont {X.-C.}\ \bibnamefont {Ma}},\ and\ \bibinfo {author} {\bibfnamefont {Q.-K.}\ \bibnamefont {Xue}},\ }\bibfield  {title} {\bibinfo {title} {Experimental observation of the quantum anomalous hall effect in a magnetic topological insulator},\ }\href {https://doi.org/10.1126/science.1234414} {\bibfield  {journal} {\bibinfo  {journal} {Science}\ }\textbf {\bibinfo {volume} {340}},\ \bibinfo {pages} {167} (\bibinfo {year} {2013})}\BibitemShut {NoStop}%
\bibitem [{\citenamefont {Sha}\ \emph {et~al.}(2024)\citenamefont {Sha}, \citenamefont {Zheng}, \citenamefont {Liu}, \citenamefont {Du}, \citenamefont {Watanabe}, \citenamefont {Taniguchi}, \citenamefont {Jia}, \citenamefont {Shi}, \citenamefont {Zhong},\ and\ \citenamefont {Chen}}]{Sha2024}%
  \BibitemOpen
  \bibfield  {author} {\bibinfo {author} {\bibfnamefont {Y.}~\bibnamefont {Sha}}, \bibinfo {author} {\bibfnamefont {J.}~\bibnamefont {Zheng}}, \bibinfo {author} {\bibfnamefont {K.}~\bibnamefont {Liu}}, \bibinfo {author} {\bibfnamefont {H.}~\bibnamefont {Du}}, \bibinfo {author} {\bibfnamefont {K.}~\bibnamefont {Watanabe}}, \bibinfo {author} {\bibfnamefont {T.}~\bibnamefont {Taniguchi}}, \bibinfo {author} {\bibfnamefont {J.}~\bibnamefont {Jia}}, \bibinfo {author} {\bibfnamefont {Z.}~\bibnamefont {Shi}}, \bibinfo {author} {\bibfnamefont {R.}~\bibnamefont {Zhong}},\ and\ \bibinfo {author} {\bibfnamefont {G.}~\bibnamefont {Chen}},\ }\bibfield  {title} {\bibinfo {title} {Observation of a chern insulator in crystalline abca-tetralayer graphene with spin-orbit coupling},\ }\href {https://doi.org/10.1126/science.adj8272} {\bibfield  {journal} {\bibinfo  {journal} {Science}\ }\textbf {\bibinfo {volume} {384}},\ \bibinfo {pages} {414} (\bibinfo {year} {2024})}\BibitemShut {NoStop}%
\bibitem [{\citenamefont {Weng}\ \emph {et~al.}(2015{\natexlab{b}})\citenamefont {Weng}, \citenamefont {Yu}, \citenamefont {Hu}, \citenamefont {Dai},\ and\ \citenamefont {Fang}}]{Weng2015b}%
  \BibitemOpen
  \bibfield  {author} {\bibinfo {author} {\bibfnamefont {H.}~\bibnamefont {Weng}}, \bibinfo {author} {\bibfnamefont {R.}~\bibnamefont {Yu}}, \bibinfo {author} {\bibfnamefont {X.}~\bibnamefont {Hu}}, \bibinfo {author} {\bibfnamefont {X.}~\bibnamefont {Dai}},\ and\ \bibinfo {author} {\bibfnamefont {Z.}~\bibnamefont {Fang}},\ }\bibfield  {title} {\bibinfo {title} {Quantum anomalous hall effect and related topological electronic states},\ }\href {https://doi.org/10.1080/00018732.2015.1068524} {\bibfield  {journal} {\bibinfo  {journal} {Advances in Physics}\ }\textbf {\bibinfo {volume} {64}},\ \bibinfo {pages} {227} (\bibinfo {year} {2015}{\natexlab{b}})}\BibitemShut {NoStop}%
\bibitem [{\citenamefont {Bai}\ \emph {et~al.}(2025)\citenamefont {Bai}, \citenamefont {Zou}, \citenamefont {Chen}, \citenamefont {Li}, \citenamefont {Yuan}, \citenamefont {Dai}, \citenamefont {Huang},\ and\ \citenamefont {Niu}}]{Bai2025}%
  \BibitemOpen
  \bibfield  {author} {\bibinfo {author} {\bibfnamefont {Y.}~\bibnamefont {Bai}}, \bibinfo {author} {\bibfnamefont {X.}~\bibnamefont {Zou}}, \bibinfo {author} {\bibfnamefont {Z.}~\bibnamefont {Chen}}, \bibinfo {author} {\bibfnamefont {R.}~\bibnamefont {Li}}, \bibinfo {author} {\bibfnamefont {B.}~\bibnamefont {Yuan}}, \bibinfo {author} {\bibfnamefont {Y.}~\bibnamefont {Dai}}, \bibinfo {author} {\bibfnamefont {B.}~\bibnamefont {Huang}},\ and\ \bibinfo {author} {\bibfnamefont {C.}~\bibnamefont {Niu}},\ }\bibfield  {title} {\bibinfo {title} {Dual chern insulators with electronic and magnonic edge states in two-dimensional ferromagnets},\ }\href {https://doi.org/10.1021/acsnano.5c00323} {\bibfield  {journal} {\bibinfo  {journal} {ACS Nano}\ }\textbf {\bibinfo {volume} {19}},\ \bibinfo {pages} {9265} (\bibinfo {year} {2025})}\BibitemShut {NoStop}%
\bibitem [{\citenamefont {Yu}\ \emph {et~al.}(2010)\citenamefont {Yu}, \citenamefont {Zhang}, \citenamefont {Zhang}, \citenamefont {Zhang}, \citenamefont {Dai},\ and\ \citenamefont {Fang}}]{Yu2010}%
  \BibitemOpen
  \bibfield  {author} {\bibinfo {author} {\bibfnamefont {R.}~\bibnamefont {Yu}}, \bibinfo {author} {\bibfnamefont {W.}~\bibnamefont {Zhang}}, \bibinfo {author} {\bibfnamefont {H.-J.}\ \bibnamefont {Zhang}}, \bibinfo {author} {\bibfnamefont {S.-C.}\ \bibnamefont {Zhang}}, \bibinfo {author} {\bibfnamefont {X.}~\bibnamefont {Dai}},\ and\ \bibinfo {author} {\bibfnamefont {Z.}~\bibnamefont {Fang}},\ }\bibfield  {title} {\bibinfo {title} {Quantized anomalous hall effect in magnetic topological insulators},\ }\href {https://doi.org/10.1126/science.1187485} {\bibfield  {journal} {\bibinfo  {journal} {Science}\ }\textbf {\bibinfo {volume} {329}},\ \bibinfo {pages} {61} (\bibinfo {year} {2010})}\BibitemShut {NoStop}%
\bibitem [{\citenamefont {Hasan}\ and\ \citenamefont {Kane}(2010)}]{Kane2010}%
  \BibitemOpen
  \bibfield  {author} {\bibinfo {author} {\bibfnamefont {M.~Z.}\ \bibnamefont {Hasan}}\ and\ \bibinfo {author} {\bibfnamefont {C.~L.}\ \bibnamefont {Kane}},\ }\bibfield  {title} {\bibinfo {title} {Colloquium: Topological insulators},\ }\href {https://doi.org/10.1103/RevModPhys.82.3045} {\bibfield  {journal} {\bibinfo  {journal} {Rev. Mod. Phys.}\ }\textbf {\bibinfo {volume} {82}},\ \bibinfo {pages} {3045} (\bibinfo {year} {2010})}\BibitemShut {NoStop}%
\bibitem [{\citenamefont {Von~Klitzing}(2005)}]{VonKlitzing2005}%
  \BibitemOpen
  \bibfield  {author} {\bibinfo {author} {\bibfnamefont {K.}~\bibnamefont {Von~Klitzing}},\ }\bibfield  {title} {\bibinfo {title} {Developments in the quantum hall effect},\ }\href {https://doi.org/10.1098/rsta.2005.1640} {\bibfield  {journal} {\bibinfo  {journal} {Philosophical Transactions of the Royal Society A: Mathematical, Physical and Engineering Sciences}\ }\textbf {\bibinfo {volume} {363}},\ \bibinfo {pages} {2203} (\bibinfo {year} {2005})}\BibitemShut {NoStop}%
\bibitem [{\citenamefont {He}\ \emph {et~al.}(2018)\citenamefont {He}, \citenamefont {Wang},\ and\ \citenamefont {Xue}}]{He2018a}%
  \BibitemOpen
  \bibfield  {author} {\bibinfo {author} {\bibfnamefont {K.}~\bibnamefont {He}}, \bibinfo {author} {\bibfnamefont {Y.}~\bibnamefont {Wang}},\ and\ \bibinfo {author} {\bibfnamefont {Q.-K.}\ \bibnamefont {Xue}},\ }\bibfield  {title} {\bibinfo {title} {Topological materials: Quantum anomalous hall system},\ }\href {https://doi.org/10.1146/annurev-conmatphys-033117-054144} {\bibfield  {journal} {\bibinfo  {journal} {Annual Review of Condensed Matter Physics}\ }\textbf {\bibinfo {volume} {9}},\ \bibinfo {pages} {329} (\bibinfo {year} {2018})}\BibitemShut {NoStop}%
\bibitem [{\citenamefont {Tokura}\ \emph {et~al.}(2019)\citenamefont {Tokura}, \citenamefont {Yasuda},\ and\ \citenamefont {Tsukazaki}}]{Tokura2019}%
  \BibitemOpen
  \bibfield  {author} {\bibinfo {author} {\bibfnamefont {Y.}~\bibnamefont {Tokura}}, \bibinfo {author} {\bibfnamefont {K.}~\bibnamefont {Yasuda}},\ and\ \bibinfo {author} {\bibfnamefont {A.}~\bibnamefont {Tsukazaki}},\ }\bibfield  {title} {\bibinfo {title} {Magnetic topological insulators},\ }\href {https://doi.org/10.1038/s42254-018-0011-5} {\bibfield  {journal} {\bibinfo  {journal} {Nature Reviews Physics}\ }\textbf {\bibinfo {volume} {1}},\ \bibinfo {pages} {126} (\bibinfo {year} {2019})}\BibitemShut {NoStop}%
\bibitem [{\citenamefont {Liang}\ \emph {et~al.}(2025)\citenamefont {Liang}, \citenamefont {Li}, \citenamefont {An}, \citenamefont {Ren}, \citenamefont {Qiao},\ and\ \citenamefont {Niu}}]{Liang2025}%
  \BibitemOpen
  \bibfield  {author} {\bibinfo {author} {\bibfnamefont {W.}~\bibnamefont {Liang}}, \bibinfo {author} {\bibfnamefont {Z.}~\bibnamefont {Li}}, \bibinfo {author} {\bibfnamefont {J.}~\bibnamefont {An}}, \bibinfo {author} {\bibfnamefont {Y.}~\bibnamefont {Ren}}, \bibinfo {author} {\bibfnamefont {Z.}~\bibnamefont {Qiao}},\ and\ \bibinfo {author} {\bibfnamefont {Q.}~\bibnamefont {Niu}},\ }\bibfield  {title} {\bibinfo {title} {Chern number tunable quantum anomalous hall effect in compensated antiferromagnets},\ }\href {https://doi.org/10.1103/PhysRevLett.134.116603} {\bibfield  {journal} {\bibinfo  {journal} {Physical Review Letters}\ }\textbf {\bibinfo {volume} {134}},\ \bibinfo {pages} {116603} (\bibinfo {year} {2025})}\BibitemShut {NoStop}%
\bibitem [{\citenamefont {Haldane}(1988)}]{Haldane1988b}%
  \BibitemOpen
  \bibfield  {author} {\bibinfo {author} {\bibfnamefont {F.~D.~M.}\ \bibnamefont {Haldane}},\ }\bibfield  {title} {\bibinfo {title} {Model for a quanum hall effect without landau levels: Condensed-matter realization of the "parity anomaly"},\ }\href {https://doi.org/10.1103/PhysRevLett.61.2015} {\bibfield  {journal} {\bibinfo  {journal} {Physical Review Letters}\ }\textbf {\bibinfo {volume} {61}},\ \bibinfo {pages} {2015} (\bibinfo {year} {1988})}\BibitemShut {NoStop}%
\bibitem [{\citenamefont {Checkelsky}\ \emph {et~al.}(2014)\citenamefont {Checkelsky}, \citenamefont {Yoshimi}, \citenamefont {Tsukazaki}, \citenamefont {Takahashi}, \citenamefont {Kozuka}, \citenamefont {Falson}, \citenamefont {Kawasaki},\ and\ \citenamefont {Tokura}}]{Checkelsky2014a}%
  \BibitemOpen
  \bibfield  {author} {\bibinfo {author} {\bibfnamefont {J.~G.}\ \bibnamefont {Checkelsky}}, \bibinfo {author} {\bibfnamefont {R.}~\bibnamefont {Yoshimi}}, \bibinfo {author} {\bibfnamefont {A.}~\bibnamefont {Tsukazaki}}, \bibinfo {author} {\bibfnamefont {K.~S.}\ \bibnamefont {Takahashi}}, \bibinfo {author} {\bibfnamefont {Y.}~\bibnamefont {Kozuka}}, \bibinfo {author} {\bibfnamefont {J.}~\bibnamefont {Falson}}, \bibinfo {author} {\bibfnamefont {M.}~\bibnamefont {Kawasaki}},\ and\ \bibinfo {author} {\bibfnamefont {Y.}~\bibnamefont {Tokura}},\ }\bibfield  {title} {\bibinfo {title} {Trajectory of the anomalous hall effect towards the quantized state in a ferromagnetic topological insulator},\ }\href {https://doi.org/10.1038/nphys3053} {\bibfield  {journal} {\bibinfo  {journal} {Nature Physics}\ }\textbf {\bibinfo {volume} {10}},\ \bibinfo {pages} {731} (\bibinfo {year} {2014})}\BibitemShut {NoStop}%
\bibitem [{\citenamefont {Mogi}\ \emph {et~al.}(2015)\citenamefont {Mogi}, \citenamefont {Yoshimi}, \citenamefont {Tsukazaki}, \citenamefont {Yasuda}, \citenamefont {Kozuka}, \citenamefont {Takahashi}, \citenamefont {Kawasaki},\ and\ \citenamefont {Tokura}}]{Mogi2015}%
  \BibitemOpen
  \bibfield  {author} {\bibinfo {author} {\bibfnamefont {M.}~\bibnamefont {Mogi}}, \bibinfo {author} {\bibfnamefont {R.}~\bibnamefont {Yoshimi}}, \bibinfo {author} {\bibfnamefont {A.}~\bibnamefont {Tsukazaki}}, \bibinfo {author} {\bibfnamefont {K.}~\bibnamefont {Yasuda}}, \bibinfo {author} {\bibfnamefont {Y.}~\bibnamefont {Kozuka}}, \bibinfo {author} {\bibfnamefont {K.~S.}\ \bibnamefont {Takahashi}}, \bibinfo {author} {\bibfnamefont {M.}~\bibnamefont {Kawasaki}},\ and\ \bibinfo {author} {\bibfnamefont {Y.}~\bibnamefont {Tokura}},\ }\bibfield  {title} {\bibinfo {title} {Magnetic modulation doping in topological insulators toward higher-temperature quantum anomalous hall effect},\ }\href {https://doi.org/10.1063/1.4935075} {\bibfield  {journal} {\bibinfo  {journal} {Applied Physics Letters}\ }\textbf {\bibinfo {volume} {107}},\ \bibinfo {pages} {182401} (\bibinfo {year} {2015})}\BibitemShut {NoStop}%
\bibitem [{\citenamefont {Deng}\ \emph {et~al.}(2020)\citenamefont {Deng}, \citenamefont {Yu}, \citenamefont {Shi}, \citenamefont {Guo}, \citenamefont {Xu}, \citenamefont {Wang}, \citenamefont {Chen},\ and\ \citenamefont {Zhang}}]{Deng2020a}%
  \BibitemOpen
  \bibfield  {author} {\bibinfo {author} {\bibfnamefont {Y.}~\bibnamefont {Deng}}, \bibinfo {author} {\bibfnamefont {Y.}~\bibnamefont {Yu}}, \bibinfo {author} {\bibfnamefont {M.~Z.}\ \bibnamefont {Shi}}, \bibinfo {author} {\bibfnamefont {Z.}~\bibnamefont {Guo}}, \bibinfo {author} {\bibfnamefont {Z.}~\bibnamefont {Xu}}, \bibinfo {author} {\bibfnamefont {J.}~\bibnamefont {Wang}}, \bibinfo {author} {\bibfnamefont {X.~H.}\ \bibnamefont {Chen}},\ and\ \bibinfo {author} {\bibfnamefont {Y.}~\bibnamefont {Zhang}},\ }\bibfield  {title} {\bibinfo {title} {Quantum anomalous hall effect in intrinsic magnetic topological insulator mnbi{\textsubscript{2}}te{\textsubscript{4}}},\ }\href {https://doi.org/10.1126/science.aax8156} {\bibfield  {journal} {\bibinfo  {journal} {Science}\ }\textbf {\bibinfo {volume} {367}},\ \bibinfo {pages} {895} (\bibinfo {year} {2020})}\BibitemShut {NoStop}%
\bibitem [{\citenamefont {Li}\ \emph {et~al.}(2019)\citenamefont {Li}, \citenamefont {Li}, \citenamefont {Du}, \citenamefont {Wang}, \citenamefont {Gu}, \citenamefont {Zhang}, \citenamefont {He}, \citenamefont {Duan},\ and\ \citenamefont {Xu}}]{Li2019b}%
  \BibitemOpen
  \bibfield  {author} {\bibinfo {author} {\bibfnamefont {J.}~\bibnamefont {Li}}, \bibinfo {author} {\bibfnamefont {Y.}~\bibnamefont {Li}}, \bibinfo {author} {\bibfnamefont {S.}~\bibnamefont {Du}}, \bibinfo {author} {\bibfnamefont {Z.}~\bibnamefont {Wang}}, \bibinfo {author} {\bibfnamefont {B.-L.}\ \bibnamefont {Gu}}, \bibinfo {author} {\bibfnamefont {S.-C.}\ \bibnamefont {Zhang}}, \bibinfo {author} {\bibfnamefont {K.}~\bibnamefont {He}}, \bibinfo {author} {\bibfnamefont {W.}~\bibnamefont {Duan}},\ and\ \bibinfo {author} {\bibfnamefont {Y.}~\bibnamefont {Xu}},\ }\bibfield  {title} {\bibinfo {title} {Intrinsic magnetic topological insulators in van der waals layered mnbi{\textsubscript{2}}te{\textsubscript{4}}-family ,materials},\ }\href {https://doi.org/10.1126/sciadv.aaw5685} {\bibfield  {journal} {\bibinfo  {journal} {Science Advances}\ }\textbf {\bibinfo {volume} {5}},\ \bibinfo {pages} {eaaw5685} (\bibinfo {year} {2019})}\BibitemShut {NoStop}%
\bibitem [{\citenamefont {Wang}\ \emph {et~al.}(2025)\citenamefont {Wang}, \citenamefont {Fu}, \citenamefont {Wang}, \citenamefont {Lian}, \citenamefont {Yang}, \citenamefont {Li}, \citenamefont {Xu}, \citenamefont {Gao}, \citenamefont {Yang}, \citenamefont {Wang}, \citenamefont {Jiang}, \citenamefont {Zhang}, \citenamefont {Wang},\ and\ \citenamefont {Liu}}]{Wang2025a}%
  \BibitemOpen
  \bibfield  {author} {\bibinfo {author} {\bibfnamefont {Y.}~\bibnamefont {Wang}}, \bibinfo {author} {\bibfnamefont {B.}~\bibnamefont {Fu}}, \bibinfo {author} {\bibfnamefont {Y.}~\bibnamefont {Wang}}, \bibinfo {author} {\bibfnamefont {Z.}~\bibnamefont {Lian}}, \bibinfo {author} {\bibfnamefont {S.}~\bibnamefont {Yang}}, \bibinfo {author} {\bibfnamefont {Y.}~\bibnamefont {Li}}, \bibinfo {author} {\bibfnamefont {L.}~\bibnamefont {Xu}}, \bibinfo {author} {\bibfnamefont {Z.}~\bibnamefont {Gao}}, \bibinfo {author} {\bibfnamefont {X.}~\bibnamefont {Yang}}, \bibinfo {author} {\bibfnamefont {W.}~\bibnamefont {Wang}}, \bibinfo {author} {\bibfnamefont {W.}~\bibnamefont {Jiang}}, \bibinfo {author} {\bibfnamefont {J.}~\bibnamefont {Zhang}}, \bibinfo {author} {\bibfnamefont {Y.}~\bibnamefont {Wang}},\ and\ \bibinfo {author} {\bibfnamefont {C.}~\bibnamefont {Liu}},\ }\bibfield  {title} {\bibinfo {title} {Towards the quantized anomalous hall effect in alox-capped mnbi{\textsubscript{2}}te{\textsubscript{4}}},\ }\href
  {https://doi.org/10.1038/s41467-025-57039-7} {\bibfield  {journal} {\bibinfo  {journal} {Nature Communications}\ }\textbf {\bibinfo {volume} {16}},\ \bibinfo {pages} {1727} (\bibinfo {year} {2025})}\BibitemShut {NoStop}%
\bibitem [{\citenamefont {Pierce}\ \emph {et~al.}(2021)\citenamefont {Pierce}, \citenamefont {Xie}, \citenamefont {Park}, \citenamefont {Khalaf}, \citenamefont {Lee}, \citenamefont {Cao}, \citenamefont {Parker}, \citenamefont {Forrester}, \citenamefont {Chen}, \citenamefont {Watanabe}, \citenamefont {Taniguchi}, \citenamefont {Vishwanath}, \citenamefont {{Jarillo-Herrero}},\ and\ \citenamefont {Yacoby}}]{Pierce2021}%
  \BibitemOpen
  \bibfield  {author} {\bibinfo {author} {\bibfnamefont {A.~T.}\ \bibnamefont {Pierce}}, \bibinfo {author} {\bibfnamefont {Y.}~\bibnamefont {Xie}}, \bibinfo {author} {\bibfnamefont {J.~M.}\ \bibnamefont {Park}}, \bibinfo {author} {\bibfnamefont {E.}~\bibnamefont {Khalaf}}, \bibinfo {author} {\bibfnamefont {S.~H.}\ \bibnamefont {Lee}}, \bibinfo {author} {\bibfnamefont {Y.}~\bibnamefont {Cao}}, \bibinfo {author} {\bibfnamefont {D.~E.}\ \bibnamefont {Parker}}, \bibinfo {author} {\bibfnamefont {P.~R.}\ \bibnamefont {Forrester}}, \bibinfo {author} {\bibfnamefont {S.}~\bibnamefont {Chen}}, \bibinfo {author} {\bibfnamefont {K.}~\bibnamefont {Watanabe}}, \bibinfo {author} {\bibfnamefont {T.}~\bibnamefont {Taniguchi}}, \bibinfo {author} {\bibfnamefont {A.}~\bibnamefont {Vishwanath}}, \bibinfo {author} {\bibfnamefont {P.}~\bibnamefont {{Jarillo-Herrero}}},\ and\ \bibinfo {author} {\bibfnamefont {A.}~\bibnamefont {Yacoby}},\ }\bibfield  {title} {\bibinfo {title} {Unconventional sequence of correlated chern insulators in
  magic-angle twisted bilayer graphene},\ }\href {https://doi.org/10.1038/s41567-021-01347-4} {\bibfield  {journal} {\bibinfo  {journal} {Nature Physics}\ }\textbf {\bibinfo {volume} {17}},\ \bibinfo {pages} {1210} (\bibinfo {year} {2021})}\BibitemShut {NoStop}%
\bibitem [{\citenamefont {Geisenhof}\ \emph {et~al.}(2021)\citenamefont {Geisenhof}, \citenamefont {Winterer}, \citenamefont {Seiler}, \citenamefont {Lenz}, \citenamefont {Xu}, \citenamefont {Zhang},\ and\ \citenamefont {Weitz}}]{Geisenhof2021}%
  \BibitemOpen
  \bibfield  {author} {\bibinfo {author} {\bibfnamefont {F.~R.}\ \bibnamefont {Geisenhof}}, \bibinfo {author} {\bibfnamefont {F.}~\bibnamefont {Winterer}}, \bibinfo {author} {\bibfnamefont {A.~M.}\ \bibnamefont {Seiler}}, \bibinfo {author} {\bibfnamefont {J.}~\bibnamefont {Lenz}}, \bibinfo {author} {\bibfnamefont {T.}~\bibnamefont {Xu}}, \bibinfo {author} {\bibfnamefont {F.}~\bibnamefont {Zhang}},\ and\ \bibinfo {author} {\bibfnamefont {R.~T.}\ \bibnamefont {Weitz}},\ }\bibfield  {title} {\bibinfo {title} {Quantum anomalous hall octet driven by orbital magnetism in bilayer graphene},\ }\href {https://doi.org/10.1038/s41586-021-03849-w} {\bibfield  {journal} {\bibinfo  {journal} {Nature}\ }\textbf {\bibinfo {volume} {598}},\ \bibinfo {pages} {53} (\bibinfo {year} {2021})}\BibitemShut {NoStop}%
\bibitem [{\citenamefont {Serlin}\ \emph {et~al.}(2020)\citenamefont {Serlin}, \citenamefont {Tschirhart}, \citenamefont {Polshyn}, \citenamefont {Zhang}, \citenamefont {Zhu}, \citenamefont {Watanabe}, \citenamefont {Taniguchi}, \citenamefont {Balents},\ and\ \citenamefont {Young}}]{Serlin2020}%
  \BibitemOpen
  \bibfield  {author} {\bibinfo {author} {\bibfnamefont {M.}~\bibnamefont {Serlin}}, \bibinfo {author} {\bibfnamefont {C.~L.}\ \bibnamefont {Tschirhart}}, \bibinfo {author} {\bibfnamefont {H.}~\bibnamefont {Polshyn}}, \bibinfo {author} {\bibfnamefont {Y.}~\bibnamefont {Zhang}}, \bibinfo {author} {\bibfnamefont {J.}~\bibnamefont {Zhu}}, \bibinfo {author} {\bibfnamefont {K.}~\bibnamefont {Watanabe}}, \bibinfo {author} {\bibfnamefont {T.}~\bibnamefont {Taniguchi}}, \bibinfo {author} {\bibfnamefont {L.}~\bibnamefont {Balents}},\ and\ \bibinfo {author} {\bibfnamefont {A.~F.}\ \bibnamefont {Young}},\ }\bibfield  {title} {\bibinfo {title} {Intrinsic quantized anomalous hall effect in a moir{\'e} heterostructure},\ }\href {https://doi.org/10.1126/science.aay5533} {\bibfield  {journal} {\bibinfo  {journal} {Science}\ }\textbf {\bibinfo {volume} {367}},\ \bibinfo {pages} {900} (\bibinfo {year} {2020})}\BibitemShut {NoStop}%
\bibitem [{\citenamefont {Zeng}\ \emph {et~al.}(2023)\citenamefont {Zeng}, \citenamefont {Xia}, \citenamefont {Kang}, \citenamefont {Zhu}, \citenamefont {Kn{\"u}ppel}, \citenamefont {Vaswani}, \citenamefont {Watanabe}, \citenamefont {Taniguchi}, \citenamefont {Mak},\ and\ \citenamefont {Shan}}]{Zeng2023}%
  \BibitemOpen
  \bibfield  {author} {\bibinfo {author} {\bibfnamefont {Y.}~\bibnamefont {Zeng}}, \bibinfo {author} {\bibfnamefont {Z.}~\bibnamefont {Xia}}, \bibinfo {author} {\bibfnamefont {K.}~\bibnamefont {Kang}}, \bibinfo {author} {\bibfnamefont {J.}~\bibnamefont {Zhu}}, \bibinfo {author} {\bibfnamefont {P.}~\bibnamefont {Kn{\"u}ppel}}, \bibinfo {author} {\bibfnamefont {C.}~\bibnamefont {Vaswani}}, \bibinfo {author} {\bibfnamefont {K.}~\bibnamefont {Watanabe}}, \bibinfo {author} {\bibfnamefont {T.}~\bibnamefont {Taniguchi}}, \bibinfo {author} {\bibfnamefont {K.~F.}\ \bibnamefont {Mak}},\ and\ \bibinfo {author} {\bibfnamefont {J.}~\bibnamefont {Shan}},\ }\bibfield  {title} {\bibinfo {title} {Thermodynamic evidence of fractional chern insulator in moir{\'e} mote{\textsubscript{2}}},\ }\href {https://doi.org/10.1038/s41586-023-06452-3} {\bibfield  {journal} {\bibinfo  {journal} {Nature}\ }\textbf {\bibinfo {volume} {622}},\ \bibinfo {pages} {69} (\bibinfo {year} {2023})}\BibitemShut {NoStop}%
\bibitem [{\citenamefont {Tao}\ \emph {et~al.}(2024)\citenamefont {Tao}, \citenamefont {Shen}, \citenamefont {Jiang}, \citenamefont {Li}, \citenamefont {Li}, \citenamefont {Ma}, \citenamefont {Zhao}, \citenamefont {Hu}, \citenamefont {Pistunova}, \citenamefont {Watanabe}, \citenamefont {Taniguchi}, \citenamefont {Heinz}, \citenamefont {Mak},\ and\ \citenamefont {Shan}}]{Tao2024}%
  \BibitemOpen
  \bibfield  {author} {\bibinfo {author} {\bibfnamefont {Z.}~\bibnamefont {Tao}}, \bibinfo {author} {\bibfnamefont {B.}~\bibnamefont {Shen}}, \bibinfo {author} {\bibfnamefont {S.}~\bibnamefont {Jiang}}, \bibinfo {author} {\bibfnamefont {T.}~\bibnamefont {Li}}, \bibinfo {author} {\bibfnamefont {L.}~\bibnamefont {Li}}, \bibinfo {author} {\bibfnamefont {L.}~\bibnamefont {Ma}}, \bibinfo {author} {\bibfnamefont {W.}~\bibnamefont {Zhao}}, \bibinfo {author} {\bibfnamefont {J.}~\bibnamefont {Hu}}, \bibinfo {author} {\bibfnamefont {K.}~\bibnamefont {Pistunova}}, \bibinfo {author} {\bibfnamefont {K.}~\bibnamefont {Watanabe}}, \bibinfo {author} {\bibfnamefont {T.}~\bibnamefont {Taniguchi}}, \bibinfo {author} {\bibfnamefont {T.~F.}\ \bibnamefont {Heinz}}, \bibinfo {author} {\bibfnamefont {K.~F.}\ \bibnamefont {Mak}},\ and\ \bibinfo {author} {\bibfnamefont {J.}~\bibnamefont {Shan}},\ }\bibfield  {title} {\bibinfo {title} {Valley-coherent quantum anomalous hall state in ab-stacked
  mote{\textsubscript{2}}/wse{\textsubscript{2}} bilayers},\ }\href {https://doi.org/10.1103/PhysRevX.14.011004} {\bibfield  {journal} {\bibinfo  {journal} {Physical Review X}\ }\textbf {\bibinfo {volume} {14}},\ \bibinfo {pages} {011004} (\bibinfo {year} {2024})}\BibitemShut {NoStop}%
\bibitem [{\citenamefont {Chang}\ \emph {et~al.}(2015)\citenamefont {Chang}, \citenamefont {Zhao}, \citenamefont {Kim}, \citenamefont {Zhang}, \citenamefont {Assaf}, \citenamefont {Heiman}, \citenamefont {Zhang}, \citenamefont {Liu}, \citenamefont {Chan},\ and\ \citenamefont {Moodera}}]{Chang2015}%
  \BibitemOpen
  \bibfield  {author} {\bibinfo {author} {\bibfnamefont {C.-Z.}\ \bibnamefont {Chang}}, \bibinfo {author} {\bibfnamefont {W.}~\bibnamefont {Zhao}}, \bibinfo {author} {\bibfnamefont {D.~Y.}\ \bibnamefont {Kim}}, \bibinfo {author} {\bibfnamefont {H.}~\bibnamefont {Zhang}}, \bibinfo {author} {\bibfnamefont {B.~A.}\ \bibnamefont {Assaf}}, \bibinfo {author} {\bibfnamefont {D.}~\bibnamefont {Heiman}}, \bibinfo {author} {\bibfnamefont {S.-C.}\ \bibnamefont {Zhang}}, \bibinfo {author} {\bibfnamefont {C.}~\bibnamefont {Liu}}, \bibinfo {author} {\bibfnamefont {M.~H.~W.}\ \bibnamefont {Chan}},\ and\ \bibinfo {author} {\bibfnamefont {J.~S.}\ \bibnamefont {Moodera}},\ }\bibfield  {title} {\bibinfo {title} {High-precision realization of robust quantum anomalous hall state in a hard ferromagnetic topological insulator},\ }\href {https://doi.org/10.1038/nmat4204} {\bibfield  {journal} {\bibinfo  {journal} {Nature Materials}\ }\textbf {\bibinfo {volume} {14}},\ \bibinfo {pages} {473} (\bibinfo {year} {2015})}\BibitemShut
  {NoStop}%
\bibitem [{\citenamefont {Kud{\l}a}\ \emph {et~al.}(2019)\citenamefont {Kud{\l}a}, \citenamefont {Dyrda{\l}}, \citenamefont {Dugaev}, \citenamefont {Berakdar},\ and\ \citenamefont {Barna{\'s}}}]{Kudla2019}%
  \BibitemOpen
  \bibfield  {author} {\bibinfo {author} {\bibfnamefont {S.}~\bibnamefont {Kud{\l}a}}, \bibinfo {author} {\bibfnamefont {A.}~\bibnamefont {Dyrda{\l}}}, \bibinfo {author} {\bibfnamefont {V.~K.}\ \bibnamefont {Dugaev}}, \bibinfo {author} {\bibfnamefont {J.}~\bibnamefont {Berakdar}},\ and\ \bibinfo {author} {\bibfnamefont {J.}~\bibnamefont {Barna{\'s}}},\ }\bibfield  {title} {\bibinfo {title} {Conduction of surface electrons in a topological insulator with spatially random magnetization},\ }\href {https://doi.org/10.1103/PhysRevB.100.205428} {\bibfield  {journal} {\bibinfo  {journal} {Physical Review B}\ }\textbf {\bibinfo {volume} {100}},\ \bibinfo {pages} {205428} (\bibinfo {year} {2019})}\BibitemShut {NoStop}%
\bibitem [{\citenamefont {Wang}\ \emph {et~al.}(2018)\citenamefont {Wang}, \citenamefont {Ou}, \citenamefont {Liu}, \citenamefont {Wang}, \citenamefont {He}, \citenamefont {Xue},\ and\ \citenamefont {Wu}}]{Wang2018}%
  \BibitemOpen
  \bibfield  {author} {\bibinfo {author} {\bibfnamefont {W.}~\bibnamefont {Wang}}, \bibinfo {author} {\bibfnamefont {Y.}~\bibnamefont {Ou}}, \bibinfo {author} {\bibfnamefont {C.}~\bibnamefont {Liu}}, \bibinfo {author} {\bibfnamefont {Y.}~\bibnamefont {Wang}}, \bibinfo {author} {\bibfnamefont {K.}~\bibnamefont {He}}, \bibinfo {author} {\bibfnamefont {Q.-K.}\ \bibnamefont {Xue}},\ and\ \bibinfo {author} {\bibfnamefont {W.}~\bibnamefont {Wu}},\ }\bibfield  {title} {\bibinfo {title} {Direct evidence of ferromagnetism in a quantum anomalous hall system},\ }\href {https://doi.org/10.1038/s41567-018-0149-1} {\bibfield  {journal} {\bibinfo  {journal} {Nature Physics}\ }\textbf {\bibinfo {volume} {14}},\ \bibinfo {pages} {791} (\bibinfo {year} {2018})}\BibitemShut {NoStop}%
\bibitem [{\citenamefont {Wang}\ \emph {et~al.}(2013)\citenamefont {Wang}, \citenamefont {Lian}, \citenamefont {Zhang}, \citenamefont {Xu},\ and\ \citenamefont {Zhang}}]{Wang2013c}%
  \BibitemOpen
  \bibfield  {author} {\bibinfo {author} {\bibfnamefont {J.}~\bibnamefont {Wang}}, \bibinfo {author} {\bibfnamefont {B.}~\bibnamefont {Lian}}, \bibinfo {author} {\bibfnamefont {H.}~\bibnamefont {Zhang}}, \bibinfo {author} {\bibfnamefont {Y.}~\bibnamefont {Xu}},\ and\ \bibinfo {author} {\bibfnamefont {S.-C.}\ \bibnamefont {Zhang}},\ }\bibfield  {title} {\bibinfo {title} {Quantum anomalous hall effect with higher plateaus},\ }\href {https://doi.org/10.1103/PhysRevLett.111.136801} {\bibfield  {journal} {\bibinfo  {journal} {Physical Review Letters}\ }\textbf {\bibinfo {volume} {111}},\ \bibinfo {pages} {136801} (\bibinfo {year} {2013})}\BibitemShut {NoStop}%
\bibitem [{\citenamefont {Zhao}\ \emph {et~al.}(2020)\citenamefont {Zhao}, \citenamefont {Zhang}, \citenamefont {Mei}, \citenamefont {Zhou}, \citenamefont {Yi}, \citenamefont {Zhang}, \citenamefont {Yu}, \citenamefont {Xiao}, \citenamefont {Wang}, \citenamefont {Samarth}, \citenamefont {Chan}, \citenamefont {Liu},\ and\ \citenamefont {Chang}}]{Zhao2020a}%
  \BibitemOpen
  \bibfield  {author} {\bibinfo {author} {\bibfnamefont {Y.-F.}\ \bibnamefont {Zhao}}, \bibinfo {author} {\bibfnamefont {R.}~\bibnamefont {Zhang}}, \bibinfo {author} {\bibfnamefont {R.}~\bibnamefont {Mei}}, \bibinfo {author} {\bibfnamefont {L.-J.}\ \bibnamefont {Zhou}}, \bibinfo {author} {\bibfnamefont {H.}~\bibnamefont {Yi}}, \bibinfo {author} {\bibfnamefont {Y.-Q.}\ \bibnamefont {Zhang}}, \bibinfo {author} {\bibfnamefont {J.}~\bibnamefont {Yu}}, \bibinfo {author} {\bibfnamefont {R.}~\bibnamefont {Xiao}}, \bibinfo {author} {\bibfnamefont {K.}~\bibnamefont {Wang}}, \bibinfo {author} {\bibfnamefont {N.}~\bibnamefont {Samarth}}, \bibinfo {author} {\bibfnamefont {M.~H.~W.}\ \bibnamefont {Chan}}, \bibinfo {author} {\bibfnamefont {C.-X.}\ \bibnamefont {Liu}},\ and\ \bibinfo {author} {\bibfnamefont {C.-Z.}\ \bibnamefont {Chang}},\ }\bibfield  {title} {\bibinfo {title} {Tuning the chern number in quantum anomalous hall insulators},\ }\href {https://doi.org/10.1038/s41586-020-3020-3} {\bibfield  {journal} {\bibinfo
  {journal} {Nature}\ }\textbf {\bibinfo {volume} {588}},\ \bibinfo {pages} {419} (\bibinfo {year} {2020})}\BibitemShut {NoStop}%
\bibitem [{\citenamefont {Guo}\ \emph {et~al.}(2024)\citenamefont {Guo}, \citenamefont {Guo},\ and\ \citenamefont {Wang}}]{Guo2024c}%
  \BibitemOpen
  \bibfield  {author} {\bibinfo {author} {\bibfnamefont {S.-D.}\ \bibnamefont {Guo}}, \bibinfo {author} {\bibfnamefont {X.-S.}\ \bibnamefont {Guo}},\ and\ \bibinfo {author} {\bibfnamefont {G.}~\bibnamefont {Wang}},\ }\bibfield  {title} {\bibinfo {title} {Valley polarization in two-dimensional tetragonal altermagnetism},\ }\href {https://doi.org/10.1103/PhysRevB.110.184408} {\bibfield  {journal} {\bibinfo  {journal} {Physical Review B}\ }\textbf {\bibinfo {volume} {110}},\ \bibinfo {pages} {184408} (\bibinfo {year} {2024})}\BibitemShut {NoStop}%
\bibitem [{\citenamefont {Liu}\ \emph {et~al.}(2021)\citenamefont {Liu}, \citenamefont {Zou}, \citenamefont {Li}, \citenamefont {Sun}, \citenamefont {Zhou},\ and\ \citenamefont {Hou}}]{Liu2021c}%
  \BibitemOpen
  \bibfield  {author} {\bibinfo {author} {\bibfnamefont {Y.}~\bibnamefont {Liu}}, \bibinfo {author} {\bibfnamefont {Z.}~\bibnamefont {Zou}}, \bibinfo {author} {\bibfnamefont {W.}~\bibnamefont {Li}}, \bibinfo {author} {\bibfnamefont {L.}~\bibnamefont {Sun}}, \bibinfo {author} {\bibfnamefont {P.}~\bibnamefont {Zhou}},\ and\ \bibinfo {author} {\bibfnamefont {P.}~\bibnamefont {Hou}},\ }\bibfield  {title} {\bibinfo {title} {Valley plarization in monolayer ferromagnetic fecl: A first-principles study},\ }\href {https://doi.org/10.1002/pssr.202000551} {\bibfield  {journal} {\bibinfo  {journal} {physica status solidi (RRL) -- Rapid Research Letters}\ }\textbf {\bibinfo {volume} {15}},\ \bibinfo {pages} {2000551} (\bibinfo {year} {2021})}\BibitemShut {NoStop}%
\bibitem [{\citenamefont {Shen}\ \emph {et~al.}(2017)\citenamefont {Shen}, \citenamefont {Tong}, \citenamefont {Gong},\ and\ \citenamefont {Duan}}]{Shen2017}%
  \BibitemOpen
  \bibfield  {author} {\bibinfo {author} {\bibfnamefont {X.-W.}\ \bibnamefont {Shen}}, \bibinfo {author} {\bibfnamefont {W.-Y.}\ \bibnamefont {Tong}}, \bibinfo {author} {\bibfnamefont {S.-J.}\ \bibnamefont {Gong}},\ and\ \bibinfo {author} {\bibfnamefont {C.-G.}\ \bibnamefont {Duan}},\ }\bibfield  {title} {\bibinfo {title} {Electrically tunable polarizer based on 2d orthorhombic ferrovalley materials},\ }\href {https://doi.org/10.1088/2053-1583/aa8d3b} {\bibfield  {journal} {\bibinfo  {journal} {2D Materials}\ }\textbf {\bibinfo {volume} {5}},\ \bibinfo {pages} {011001} (\bibinfo {year} {2017})}\BibitemShut {NoStop}%
\bibitem [{\citenamefont {Chen}\ \emph {et~al.}(2025)\citenamefont {Chen}, \citenamefont {Chen}, \citenamefont {Cheng}, \citenamefont {Zhao}, \citenamefont {Hu}, \citenamefont {Yuan},\ and\ \citenamefont {Ren}}]{Ren2025}%
  \BibitemOpen
  \bibfield  {author} {\bibinfo {author} {\bibfnamefont {H.}~\bibnamefont {Chen}}, \bibinfo {author} {\bibfnamefont {F.}~\bibnamefont {Chen}}, \bibinfo {author} {\bibfnamefont {H.}~\bibnamefont {Cheng}}, \bibinfo {author} {\bibfnamefont {X.}~\bibnamefont {Zhao}}, \bibinfo {author} {\bibfnamefont {G.}~\bibnamefont {Hu}}, \bibinfo {author} {\bibfnamefont {X.}~\bibnamefont {Yuan}},\ and\ \bibinfo {author} {\bibfnamefont {J.}~\bibnamefont {Ren}},\ }\bibfield  {title} {\bibinfo {title} {Layer-locked multiple valley hall effects in tetragonal altermagnetic/ferromagnetic monolayers ${M}_{2}\mathrm{SiC}{X}_{2}$ $(m=\text{transition metal}; x=\mathrm{S}, \mathrm{Se})$},\ }\href {https://doi.org/10.1103/PhysRevB.111.155428} {\bibfield  {journal} {\bibinfo  {journal} {Phys. Rev. B}\ }\textbf {\bibinfo {volume} {111}},\ \bibinfo {pages} {155428} (\bibinfo {year} {2025})}\BibitemShut {NoStop}%
\bibitem [{\citenamefont {Giannozzi}\ \emph {et~al.}(2009)\citenamefont {Giannozzi}, \citenamefont {Baroni}, \citenamefont {Bonini}, \citenamefont {Calandra}, \citenamefont {Car}, \citenamefont {Cavazzoni}, \citenamefont {Ceresoli}, \citenamefont {Chiarotti}, \citenamefont {Cococcioni}, \citenamefont {Dabo}, \citenamefont {Dal~Corso}, \citenamefont {De~Gironcoli}, \citenamefont {Fabris}, \citenamefont {Fratesi}, \citenamefont {Gebauer}, \citenamefont {Gerstmann}, \citenamefont {Gougoussis}, \citenamefont {Kokalj}, \citenamefont {Lazzeri}, \citenamefont {{Martin-Samos}}, \citenamefont {Marzari}, \citenamefont {Mauri}, \citenamefont {Mazzarello}, \citenamefont {Paolini}, \citenamefont {Pasquarello}, \citenamefont {Paulatto}, \citenamefont {Sbraccia}, \citenamefont {Scandolo}, \citenamefont {Sclauzero}, \citenamefont {Seitsonen}, \citenamefont {Smogunov}, \citenamefont {Umari},\ and\ \citenamefont {Wentzcovitch}}]{Giannozzi2009}%
  \BibitemOpen
  \bibfield  {author} {\bibinfo {author} {\bibfnamefont {P.}~\bibnamefont {Giannozzi}}, \bibinfo {author} {\bibfnamefont {S.}~\bibnamefont {Baroni}}, \bibinfo {author} {\bibfnamefont {N.}~\bibnamefont {Bonini}}, \bibinfo {author} {\bibfnamefont {M.}~\bibnamefont {Calandra}}, \bibinfo {author} {\bibfnamefont {R.}~\bibnamefont {Car}}, \bibinfo {author} {\bibfnamefont {C.}~\bibnamefont {Cavazzoni}}, \bibinfo {author} {\bibfnamefont {D.}~\bibnamefont {Ceresoli}}, \bibinfo {author} {\bibfnamefont {G.~L.}\ \bibnamefont {Chiarotti}}, \bibinfo {author} {\bibfnamefont {M.}~\bibnamefont {Cococcioni}}, \bibinfo {author} {\bibfnamefont {I.}~\bibnamefont {Dabo}}, \bibinfo {author} {\bibfnamefont {A.}~\bibnamefont {Dal~Corso}}, \bibinfo {author} {\bibfnamefont {S.}~\bibnamefont {De~Gironcoli}}, \bibinfo {author} {\bibfnamefont {S.}~\bibnamefont {Fabris}}, \bibinfo {author} {\bibfnamefont {G.}~\bibnamefont {Fratesi}}, \bibinfo {author} {\bibfnamefont {R.}~\bibnamefont {Gebauer}}, \bibinfo {author} {\bibfnamefont
  {U.}~\bibnamefont {Gerstmann}}, \bibinfo {author} {\bibfnamefont {C.}~\bibnamefont {Gougoussis}}, \bibinfo {author} {\bibfnamefont {A.}~\bibnamefont {Kokalj}}, \bibinfo {author} {\bibfnamefont {M.}~\bibnamefont {Lazzeri}}, \bibinfo {author} {\bibfnamefont {L.}~\bibnamefont {{Martin-Samos}}}, \bibinfo {author} {\bibfnamefont {N.}~\bibnamefont {Marzari}}, \bibinfo {author} {\bibfnamefont {F.}~\bibnamefont {Mauri}}, \bibinfo {author} {\bibfnamefont {R.}~\bibnamefont {Mazzarello}}, \bibinfo {author} {\bibfnamefont {S.}~\bibnamefont {Paolini}}, \bibinfo {author} {\bibfnamefont {A.}~\bibnamefont {Pasquarello}}, \bibinfo {author} {\bibfnamefont {L.}~\bibnamefont {Paulatto}}, \bibinfo {author} {\bibfnamefont {C.}~\bibnamefont {Sbraccia}}, \bibinfo {author} {\bibfnamefont {S.}~\bibnamefont {Scandolo}}, \bibinfo {author} {\bibfnamefont {G.}~\bibnamefont {Sclauzero}}, \bibinfo {author} {\bibfnamefont {A.~P.}\ \bibnamefont {Seitsonen}}, \bibinfo {author} {\bibfnamefont {A.}~\bibnamefont {Smogunov}}, \bibinfo {author}
  {\bibfnamefont {P.}~\bibnamefont {Umari}},\ and\ \bibinfo {author} {\bibfnamefont {R.~M.}\ \bibnamefont {Wentzcovitch}},\ }\bibfield  {title} {\bibinfo {title} {Quantum espresso: A modular and open-source software project for quantum simulations of materials},\ }\href {https://doi.org/10.1088/0953-8984/21/39/395502} {\bibfield  {journal} {\bibinfo  {journal} {Journal of Physics: Condensed Matter}\ }\textbf {\bibinfo {volume} {21}},\ \bibinfo {pages} {395502} (\bibinfo {year} {2009})}\BibitemShut {NoStop}%
\bibitem [{\citenamefont {Giannozzi}\ \emph {et~al.}(2017)\citenamefont {Giannozzi}, \citenamefont {Andreussi}, \citenamefont {Brumme}, \citenamefont {Bunau}, \citenamefont {Buongiorno~Nardelli}, \citenamefont {Calandra}, \citenamefont {Car}, \citenamefont {Cavazzoni}, \citenamefont {Ceresoli}, \citenamefont {Cococcioni}, \citenamefont {Colonna}, \citenamefont {Carnimeo}, \citenamefont {Dal~Corso}, \citenamefont {De~Gironcoli}, \citenamefont {Delugas}, \citenamefont {DiStasio}, \citenamefont {Ferretti}, \citenamefont {Floris}, \citenamefont {Fratesi}, \citenamefont {Fugallo}, \citenamefont {Gebauer}, \citenamefont {Gerstmann}, \citenamefont {Giustino}, \citenamefont {Gorni}, \citenamefont {Jia}, \citenamefont {Kawamura}, \citenamefont {Ko}, \citenamefont {Kokalj}, \citenamefont {K{\"u}{\c c}{\"u}kbenli}, \citenamefont {Lazzeri}, \citenamefont {Marsili}, \citenamefont {Marzari}, \citenamefont {Mauri}, \citenamefont {Nguyen}, \citenamefont {Nguyen}, \citenamefont {{Otero-de-la-Roza}}, \citenamefont {Paulatto},
  \citenamefont {Ponc{\'e}}, \citenamefont {Rocca}, \citenamefont {Sabatini}, \citenamefont {Santra}, \citenamefont {Schlipf}, \citenamefont {Seitsonen}, \citenamefont {Smogunov}, \citenamefont {Timrov}, \citenamefont {Thonhauser}, \citenamefont {Umari}, \citenamefont {Vast}, \citenamefont {Wu},\ and\ \citenamefont {Baroni}}]{Giannozzi2017}%
  \BibitemOpen
  \bibfield  {author} {\bibinfo {author} {\bibfnamefont {P.}~\bibnamefont {Giannozzi}}, \bibinfo {author} {\bibfnamefont {O.}~\bibnamefont {Andreussi}}, \bibinfo {author} {\bibfnamefont {T.}~\bibnamefont {Brumme}}, \bibinfo {author} {\bibfnamefont {O.}~\bibnamefont {Bunau}}, \bibinfo {author} {\bibfnamefont {M.}~\bibnamefont {Buongiorno~Nardelli}}, \bibinfo {author} {\bibfnamefont {M.}~\bibnamefont {Calandra}}, \bibinfo {author} {\bibfnamefont {R.}~\bibnamefont {Car}}, \bibinfo {author} {\bibfnamefont {C.}~\bibnamefont {Cavazzoni}}, \bibinfo {author} {\bibfnamefont {D.}~\bibnamefont {Ceresoli}}, \bibinfo {author} {\bibfnamefont {M.}~\bibnamefont {Cococcioni}}, \bibinfo {author} {\bibfnamefont {N.}~\bibnamefont {Colonna}}, \bibinfo {author} {\bibfnamefont {I.}~\bibnamefont {Carnimeo}}, \bibinfo {author} {\bibfnamefont {A.}~\bibnamefont {Dal~Corso}}, \bibinfo {author} {\bibfnamefont {S.}~\bibnamefont {De~Gironcoli}}, \bibinfo {author} {\bibfnamefont {P.}~\bibnamefont {Delugas}}, \bibinfo {author} {\bibfnamefont
  {R.~A.}\ \bibnamefont {DiStasio}}, \bibinfo {author} {\bibfnamefont {A.}~\bibnamefont {Ferretti}}, \bibinfo {author} {\bibfnamefont {A.}~\bibnamefont {Floris}}, \bibinfo {author} {\bibfnamefont {G.}~\bibnamefont {Fratesi}}, \bibinfo {author} {\bibfnamefont {G.}~\bibnamefont {Fugallo}}, \bibinfo {author} {\bibfnamefont {R.}~\bibnamefont {Gebauer}}, \bibinfo {author} {\bibfnamefont {U.}~\bibnamefont {Gerstmann}}, \bibinfo {author} {\bibfnamefont {F.}~\bibnamefont {Giustino}}, \bibinfo {author} {\bibfnamefont {T.}~\bibnamefont {Gorni}}, \bibinfo {author} {\bibfnamefont {J.}~\bibnamefont {Jia}}, \bibinfo {author} {\bibfnamefont {M.}~\bibnamefont {Kawamura}}, \bibinfo {author} {\bibfnamefont {H.-Y.}\ \bibnamefont {Ko}}, \bibinfo {author} {\bibfnamefont {A.}~\bibnamefont {Kokalj}}, \bibinfo {author} {\bibfnamefont {E.}~\bibnamefont {K{\"u}{\c c}{\"u}kbenli}}, \bibinfo {author} {\bibfnamefont {M.}~\bibnamefont {Lazzeri}}, \bibinfo {author} {\bibfnamefont {M.}~\bibnamefont {Marsili}}, \bibinfo {author}
  {\bibfnamefont {N.}~\bibnamefont {Marzari}}, \bibinfo {author} {\bibfnamefont {F.}~\bibnamefont {Mauri}}, \bibinfo {author} {\bibfnamefont {N.~L.}\ \bibnamefont {Nguyen}}, \bibinfo {author} {\bibfnamefont {H.-V.}\ \bibnamefont {Nguyen}}, \bibinfo {author} {\bibfnamefont {A.}~\bibnamefont {{Otero-de-la-Roza}}}, \bibinfo {author} {\bibfnamefont {L.}~\bibnamefont {Paulatto}}, \bibinfo {author} {\bibfnamefont {S.}~\bibnamefont {Ponc{\'e}}}, \bibinfo {author} {\bibfnamefont {D.}~\bibnamefont {Rocca}}, \bibinfo {author} {\bibfnamefont {R.}~\bibnamefont {Sabatini}}, \bibinfo {author} {\bibfnamefont {B.}~\bibnamefont {Santra}}, \bibinfo {author} {\bibfnamefont {M.}~\bibnamefont {Schlipf}}, \bibinfo {author} {\bibfnamefont {A.~P.}\ \bibnamefont {Seitsonen}}, \bibinfo {author} {\bibfnamefont {A.}~\bibnamefont {Smogunov}}, \bibinfo {author} {\bibfnamefont {I.}~\bibnamefont {Timrov}}, \bibinfo {author} {\bibfnamefont {T.}~\bibnamefont {Thonhauser}}, \bibinfo {author} {\bibfnamefont {P.}~\bibnamefont {Umari}}, \bibinfo
  {author} {\bibfnamefont {N.}~\bibnamefont {Vast}}, \bibinfo {author} {\bibfnamefont {X.}~\bibnamefont {Wu}},\ and\ \bibinfo {author} {\bibfnamefont {S.}~\bibnamefont {Baroni}},\ }\bibfield  {title} {\bibinfo {title} {Advanced capabilities for materials modelling with quantum espresso},\ }\href {https://doi.org/10.1088/1361-648X/aa8f79} {\bibfield  {journal} {\bibinfo  {journal} {Journal of Physics: Condensed Matter}\ }\textbf {\bibinfo {volume} {29}},\ \bibinfo {pages} {465901} (\bibinfo {year} {2017})}\BibitemShut {NoStop}%
\bibitem [{\citenamefont {Perdew}\ \emph {et~al.}(1996)\citenamefont {Perdew}, \citenamefont {Burke},\ and\ \citenamefont {Ernzerhof}}]{Perdew1996}%
  \BibitemOpen
  \bibfield  {author} {\bibinfo {author} {\bibfnamefont {J.~P.}\ \bibnamefont {Perdew}}, \bibinfo {author} {\bibfnamefont {K.}~\bibnamefont {Burke}},\ and\ \bibinfo {author} {\bibfnamefont {M.}~\bibnamefont {Ernzerhof}},\ }\bibfield  {title} {\bibinfo {title} {Generalized gradient approximation made simple},\ }\href {https://doi.org/10.1103/PhysRevLett.77.3865} {\bibfield  {journal} {\bibinfo  {journal} {Physical Review Letters}\ }\textbf {\bibinfo {volume} {77}},\ \bibinfo {pages} {3865} (\bibinfo {year} {1996})}\BibitemShut {NoStop}%
\bibitem [{\citenamefont {Cococcioni}\ and\ \citenamefont {De~Gironcoli}(2005)}]{Cococcioni2005}%
  \BibitemOpen
  \bibfield  {author} {\bibinfo {author} {\bibfnamefont {M.}~\bibnamefont {Cococcioni}}\ and\ \bibinfo {author} {\bibfnamefont {S.}~\bibnamefont {De~Gironcoli}},\ }\bibfield  {title} {\bibinfo {title} {Linear response approach to the calculation of the effective interaction parameters in the lda + u method},\ }\href {https://doi.org/10.1103/PhysRevB.71.035105} {\bibfield  {journal} {\bibinfo  {journal} {Physical Review B}\ }\textbf {\bibinfo {volume} {71}},\ \bibinfo {pages} {035105} (\bibinfo {year} {2005})}\BibitemShut {NoStop}%
\bibitem [{\citenamefont {Liechtenstein}\ \emph {et~al.}(1995)\citenamefont {Liechtenstein}, \citenamefont {Anisimov},\ and\ \citenamefont {Zaanen}}]{Liechtenstein1995}%
  \BibitemOpen
  \bibfield  {author} {\bibinfo {author} {\bibfnamefont {A.~I.}\ \bibnamefont {Liechtenstein}}, \bibinfo {author} {\bibfnamefont {V.~I.}\ \bibnamefont {Anisimov}},\ and\ \bibinfo {author} {\bibfnamefont {J.}~\bibnamefont {Zaanen}},\ }\bibfield  {title} {\bibinfo {title} {Density-functional theory and strong interactions: Orbital ordering in mott-hubbard insulators},\ }\href {https://doi.org/10.1103/PhysRevB.52.R5467} {\bibfield  {journal} {\bibinfo  {journal} {Physical Review B}\ }\textbf {\bibinfo {volume} {52}},\ \bibinfo {pages} {R5467} (\bibinfo {year} {1995})}\BibitemShut {NoStop}%
\bibitem [{\citenamefont {Methfessel}\ and\ \citenamefont {Paxton}(1989)}]{Methfessel1989}%
  \BibitemOpen
  \bibfield  {author} {\bibinfo {author} {\bibfnamefont {M.}~\bibnamefont {Methfessel}}\ and\ \bibinfo {author} {\bibfnamefont {A.~T.}\ \bibnamefont {Paxton}},\ }\bibfield  {title} {\bibinfo {title} {High-precision sampling for brillouin-zone integration in metals},\ }\href {https://doi.org/10.1103/PhysRevB.40.3616} {\bibfield  {journal} {\bibinfo  {journal} {Physical Review B}\ }\textbf {\bibinfo {volume} {40}},\ \bibinfo {pages} {3616} (\bibinfo {year} {1989})}\BibitemShut {NoStop}%
\bibitem [{\citenamefont {Togo}\ and\ \citenamefont {Tanaka}(2015)}]{Togo2015}%
  \BibitemOpen
  \bibfield  {author} {\bibinfo {author} {\bibfnamefont {A.}~\bibnamefont {Togo}}\ and\ \bibinfo {author} {\bibfnamefont {I.}~\bibnamefont {Tanaka}},\ }\bibfield  {title} {\bibinfo {title} {First principles phonon calculations in materials science},\ }\href {https://doi.org/10.1016/j.scriptamat.2015.07.021} {\bibfield  {journal} {\bibinfo  {journal} {Scripta Materialia}\ }\textbf {\bibinfo {volume} {108}},\ \bibinfo {pages} {1} (\bibinfo {year} {2015})}\BibitemShut {NoStop}%
\bibitem [{\citenamefont {Zhang}\ \emph {et~al.}(2021)\citenamefont {Zhang}, \citenamefont {Wang}, \citenamefont {Guo}, \citenamefont {Li},\ and\ \citenamefont {Wang}}]{Zhang2021}%
  \BibitemOpen
  \bibfield  {author} {\bibinfo {author} {\bibfnamefont {Y.}~\bibnamefont {Zhang}}, \bibinfo {author} {\bibfnamefont {B.}~\bibnamefont {Wang}}, \bibinfo {author} {\bibfnamefont {Y.}~\bibnamefont {Guo}}, \bibinfo {author} {\bibfnamefont {Q.}~\bibnamefont {Li}},\ and\ \bibinfo {author} {\bibfnamefont {J.}~\bibnamefont {Wang}},\ }\bibfield  {title} {\bibinfo {title} {A universal framework for metropolis monte carlo simulation of magnetic curie temperature},\ }\href {https://doi.org/10.1016/j.commatsci.2021.110638} {\bibfield  {journal} {\bibinfo  {journal} {Computational Materials Science}\ }\textbf {\bibinfo {volume} {197}},\ \bibinfo {pages} {110638} (\bibinfo {year} {2021})}\BibitemShut {NoStop}%
\bibitem [{\citenamefont {Mostofi}\ \emph {et~al.}(2008)\citenamefont {Mostofi}, \citenamefont {Yates}, \citenamefont {Lee}, \citenamefont {Souza}, \citenamefont {Vanderbilt},\ and\ \citenamefont {Marzari}}]{Mostofi2008}%
  \BibitemOpen
  \bibfield  {author} {\bibinfo {author} {\bibfnamefont {A.~A.}\ \bibnamefont {Mostofi}}, \bibinfo {author} {\bibfnamefont {J.~R.}\ \bibnamefont {Yates}}, \bibinfo {author} {\bibfnamefont {Y.-S.}\ \bibnamefont {Lee}}, \bibinfo {author} {\bibfnamefont {I.}~\bibnamefont {Souza}}, \bibinfo {author} {\bibfnamefont {D.}~\bibnamefont {Vanderbilt}},\ and\ \bibinfo {author} {\bibfnamefont {N.}~\bibnamefont {Marzari}},\ }\bibfield  {title} {\bibinfo {title} {Wannier90: A tool for obtaining maximally-localised wannier functions},\ }\href {https://doi.org/10.1016/j.cpc.2007.11.016} {\bibfield  {journal} {\bibinfo  {journal} {Computer Physics Communications}\ }\textbf {\bibinfo {volume} {178}},\ \bibinfo {pages} {685} (\bibinfo {year} {2008})}\BibitemShut {NoStop}%
\bibitem [{\citenamefont {Wu}\ \emph {et~al.}(2018)\citenamefont {Wu}, \citenamefont {Zhang}, \citenamefont {Song}, \citenamefont {Troyer},\ and\ \citenamefont {Soluyanov}}]{Wu2018}%
  \BibitemOpen
  \bibfield  {author} {\bibinfo {author} {\bibfnamefont {Q.}~\bibnamefont {Wu}}, \bibinfo {author} {\bibfnamefont {S.}~\bibnamefont {Zhang}}, \bibinfo {author} {\bibfnamefont {H.-F.}\ \bibnamefont {Song}}, \bibinfo {author} {\bibfnamefont {M.}~\bibnamefont {Troyer}},\ and\ \bibinfo {author} {\bibfnamefont {A.~A.}\ \bibnamefont {Soluyanov}},\ }\bibfield  {title} {\bibinfo {title} {Wanniertools: An open-source software package for novel topological materials},\ }\href {https://doi.org/10.1016/j.cpc.2017.09.033} {\bibfield  {journal} {\bibinfo  {journal} {Computer Physics Communications}\ }\textbf {\bibinfo {volume} {224}},\ \bibinfo {pages} {405} (\bibinfo {year} {2018})}\BibitemShut {NoStop}%
\bibitem [{\citenamefont {Stokes}\ and\ \citenamefont {Hatch}(2005)}]{findsym}%
  \BibitemOpen
  \bibfield  {author} {\bibinfo {author} {\bibfnamefont {H.~T.}\ \bibnamefont {Stokes}}\ and\ \bibinfo {author} {\bibfnamefont {D.~M.}\ \bibnamefont {Hatch}},\ }\bibfield  {title} {\bibinfo {title} {Findsym: Program for identifying the space-group symmetry of a crystal},\ }\href@noop {} {\bibfield  {journal} {\bibinfo  {journal} {Applied Crystallography}\ }\textbf {\bibinfo {volume} {38}},\ \bibinfo {pages} {237} (\bibinfo {year} {2005})}\BibitemShut {NoStop}%
\bibitem [{\citenamefont {Stokes}\ \emph {et~al.}(2005)\citenamefont {Stokes}, \citenamefont {Hatch},\ and\ \citenamefont {Campbell}}]{Stokes_FINDSYM}%
  \BibitemOpen
  \bibfield  {author} {\bibinfo {author} {\bibfnamefont {H.~T.}\ \bibnamefont {Stokes}}, \bibinfo {author} {\bibfnamefont {D.~M.}\ \bibnamefont {Hatch}},\ and\ \bibinfo {author} {\bibfnamefont {B.~J.}\ \bibnamefont {Campbell}},\ }\href {http://iso.byu.edu} {\bibinfo {title} {Findsym}},\ \bibinfo {howpublished} {ISOTROPY Software Suite} (\bibinfo {year} {2005}),\ \bibinfo {note} {computer program for identifying the space group symmetry of a crystal}\BibitemShut {NoStop}%
\end{thebibliography}%

\end{document}